\documentclass[fleqn,usenatbib]{mnras}

\usepackage{newtxtext,newtxmath}

\usepackage[T1]{fontenc}

\DeclareRobustCommand{\VAN}[3]{#2}
\let\VANthebibliography\thebibliography
\def\thebibliography{\DeclareRobustCommand{\VAN}[3]{##3}\VANthebibliography}

\usepackage{graphicx}	
\usepackage{amsmath}	
\usepackage{amssymb}	
\usepackage[capitalise]{cleveref}		
\usepackage{booktabs}		
\usepackage[dvipsnames]{xcolor}   
\usepackage{bm}				
\usepackage{soul}			
\usepackage{comment}
\numberwithin{equation}{section}	
\usepackage{array} 

\newcommand{\der}{\text{d}} 

\title[Double-lens Scintillometry]{
Double-lens Scintillometry: The variable scintillation of pulsar B1508+55
}

\author[T. Sprenger et al.]{
Tim Sprenger,$^{1}$\thanks{E-mail: tsprenger@mpifr-bonn.mpg.de}
Robert Main,$^{1}$
Olaf Wucknitz,$^{1}$
Geetam Mall,$^{1,2}$
Jason Wu$^{1}$
\\
$^{1}$Max-Planck-Institut f\"ur Radioastronomie, Auf dem H\"ugel 69, 53121 Bonn, Germany
\\
$^{2}$Canadian Institute for Theoretical Astrophysics, University of Toronto, 60 St. George Street, Toronto, ON M5S 3H8, Canada \\
}

\date{Accepted XXX. Received YYY; in original form ZZZ}

\pubyear{2022}

\begin{document}
\label{firstpage}
\pagerange{\pageref{firstpage}--\pageref{lastpage}}
\maketitle

\begin{abstract}

We report on observations of PSR B1508+55's scintillation at the Effelsberg 100-m telescope spanning from early 2020 to early 2022. In the autumn of 2020, close to the time the pulsar was predicted to cross echoes in its pulse profile, a sudden transition in the scintillation arcs from peculiar stripe-like features to parabolic arclets was observed. To infer a geometric model of the scattering we measure the effects of the annual velocity curve of Earth, of the relative movement of the line of sight, and of the projection of points on a second scattering screen and develop novel methods to do so. The latter phenomenon was discovered by this study and strongly indicates a two-screen scattering geometry. We derive an analytical two-screen model and demonstrate in a Markov Chain Monte Carlo analysis as well as simulations that it can be successfully applied to explain the observations by interpreting the transition as a change of relative amplitudes of images as well as a shift in the orientation of anisotropy. The collection of methods we demonstrate here is transferable to other pulsars with the potential to strongly improve constraints on scattering models.

\end{abstract}




\section{Introduction}
\label{Sec:Introduction}

Compact radio sources like pulsars, fast radio bursts (FRBs), and a few quasars emit spatially coherent radiation which can be scattered by the ionized interstellar medium (IISM). Interference of the scattered paths of propagation leads to a variation of the intensity of the measured signal that is called scintillation. The characteristic time and frequency scales of this variation depend on the geometrical situation. In particular, they depend on the proper motion of the source perpendicular to the line of sight. The science of using scintillation to determine astrometric properties like orbits of pulsars and the distances of scattering structures within the IISM has become known as \textit{scintillometry}, while imaging these scattering structures can be understood as a form of in-line holography \citep{2008MNRAS.388.1214W}.

Interstellar scintillation (ISS) can be observed as intensity variations on many different scales. The variations studied here manifest in intensity variations within the timespans and bandwidths of single observations that tend not to alter the mean flux of the source. This kind of scintillation is caused by the Earth moving through an interference pattern. There are also long-time intensity variations. These are an incoherent effect that is caused by relative movement such that different parts of the scattering material cross the line of sight. Thus, the amplification of the flux changes.

Scintillation resembles an interference pattern in the dynamic spectrum over time and frequency. Its structure is often more clear in the power spectrum of the dynamic spectrum, which is called the seconday spectrum. Secondary spectra are dominated by parabolic structures refered to as scintillation arcs. These arcs were first described in \citet{2001ApJ...549L..97S}. \citet{2004MNRAS.354...43W,2006ApJ...637..346C} successfully explained them by invoking thin and highly anisotropic scattering screens whose interstellar optical properties were analysed by \citet{1998ApJ...505..928G}.

The physical nature of these screens, that are solely observable through the phase delay they impose on passing radiation, still needs to be determined on the small angular scales that cause scattering. Supernova remnants (SNR) are very likely physical associations, as shown for the Crab pulsar \citep{2004ApJ...612..375C}, J0538+2817 \citep{2021NatAs...5..788Y}, and likely for J0540$-$6919 \citep{2021MNRAS.505.4468G}. Other likely associations are hot stars \citep[e.g.][]{2017ApJ...843...15W} and extended HII regions \citep[e.g.][]{2022MNRAS.511.1104M}. Scattering is caused by variable densities of dispersive material that delays the arrival of pulses. Thus, scattering screens also cause variations in dispersion which is quantified by the dispersion measure (DM) of a pulsar. This correlation has been observed \citep[e.g.][]{2018MNRAS.479.4216M}.

On the theoretical side, \citet{2014MNRAS.442.3338P,2018MNRAS.478..983S} proposed folded sheets of scattering material where anisotropy is a result of inclination and that could resemble corrugated reconnection sheets of plasma at the boundary between magnetic field configurations. \citet{2019MNRAS.486.2809G,2019MNRAS.489.3692G} proposed magnetic noodles of plasma whose anisotropy is maintained by parallel magnetic fields.

\begin{figure}
 \includegraphics[width=\columnwidth]{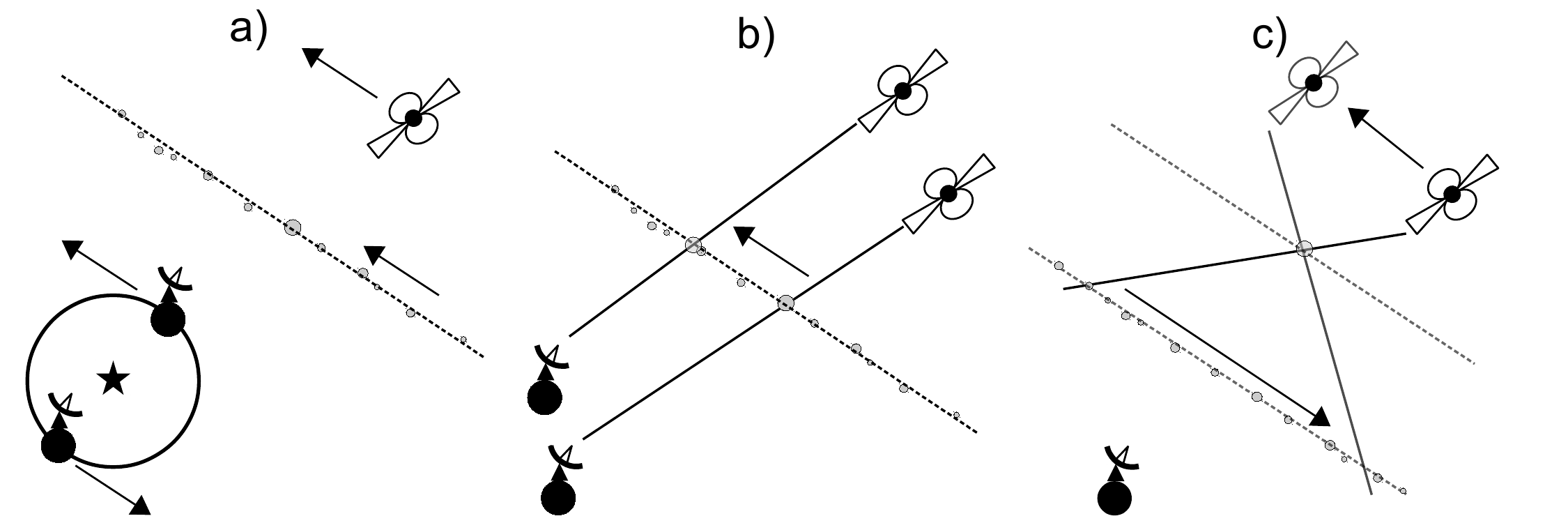}
 \vspace{-3mm}
 \caption{a) The width of scintillation arcs depends on the distances and velocities within the system and varies because of the Earth's motion around the Sun. b) The line of sight to the pulsar moves such that the positions of fixed features in the scattering structure change with respect to it. c) If there is a second scattering screen, its projected magnification moves along the scattering structures of the first screen.}
 \label{Fig:Sketches}
\end{figure}

This paper is organized as follows. An introduction to scintillation and PSR B1508+55 is given in \cref{Sec:Introduction}. The data is described in \cref{Sec:data} and \cref{Sec:Phenomenology} gives an overview of the observed phenomena. \Cref{Sec:Scint_Arcs,Sec:Feature_Alignment,Sec:Eigenvectors} contain the three methods used to extract information from the data, and whose physical interpretation is visualized in \cref{Fig:Sketches}. To explain these, a two-screen model is introduced in \cref{Sec:Two_Screens}. Computational methods are described in \cref{Sec:Computation} and the results based on them are presented in \cref{Sec:Results}. We discuss ramifications of our work, and future avenues for research in \cref{Sec:Conclusions}.

\subsection{Definitions and basic Theory}

\begin{table}
	\centering
	\begin{tabular}{>{\raggedright\arraybackslash}m{1.5cm}m{6.0cm}}
     \hline
     {\bf Symbol} & {\bf Description} \\
     \hline 
     \multicolumn{2}{c}{\it Quantities}\\
     \hline
     $\eta$ & parabolic curvature \\
     $\partial_t\sqrt{\tau}$ & $=(2\nu\sqrt{\eta})^{-1}$, effective drift rate \\
     $D_\text{eff}$ & effective distance of the lens \\
     $d_\text{scr}$ & distance of screen \\
     $d_\text{psr}$ & distance of the pulsar \\
     $\mu_\text{RA}, \mu_\text{Dec}$ & angular proper motion (right ascension, declination) \\
     $\bm{V}_\text{eff}$ & effective velocity \\
     $\bm{V}_\text{ISS}$ & scintillation speed  \\
     $\bm{V}_\oplus$ & velocity of Earth/observer \\
     $\bm{V}_\text{scr}$ & velocity of screen \\
     $\bm{V}_\text{psr}$ & velocity of the pulsar \\
     $\bm{\theta}_0$ & angular position of the central path through the screen \\
     $N_t,N_\nu,N_\theta$ & number of numerical bins (time, frequency, angle) \\
     $V_m$ & modulation speed \\
     $A_{\shortparallel},A_{\perp}$ & projection of a vector $\bm{A}$ onto a screen's axis of anisotropy (parallel, perpendicular) \\
     \hline 
     \multicolumn{2}{c}{\it Coordinates}\\
     \hline
     $t$ & time \\
     $\nu$ & frequency \\
     $f_\text{D}$ & Doppler rate (conjugate time) \\
     $\tau$ & delay (conjugate frequency) \\
     $\bm{\theta},\theta$ & angular position \\
     \hline 
     \multicolumn{2}{c}{\it Fields}\\
     \hline
     $E(t,\nu)$ & electric field \\
     $I(t,\nu)$ & dynamic spectrum \\
     $S(f_\text{D},\tau)$ & secondary spectrum \\
     $\mu(\theta)$ & complex amplitude distribution of images \\
	 \hline
	\end{tabular}
	\caption{Glossary of used variables. All vectors are two-dimensional because the dimension along the line of sight has a negligible impact on scintillation. An extended notation is used for the two-screen theory in \cref{Sec:Two_Screens,Sec:Computation,Sec:Results} and defined in \cref{Sec:Derivations} and \cref{Fig:DoubleScreenVisualization,Fig:DoubleScreenNotation}.}
	\label{tab:glossary}
\end{table}

A glossary of the variables used is given in \cref{tab:glossary}.

Scintillation inflicts a complex amplification on the observed electric field  $E(t,\nu)$. Since the incoming radiation is not temporally coherent, its phase contains no usable information. Hence, its square modulus $I(t,\nu)=\vert E \vert^2(t,\nu)$ -- the dynamic spectrum -- is used. The dynamic spectrum still contains information that is not originating from scintillation in form of the intrinsic amplitude of the signal. As a result, the mean amplitude cannot be used for this analysis. Furthermore, intrinsic variations, such as in particular pulse-to-pulse variations, present a source of noise on the scintillation signal.

The secondary spectrum $S(f_\text{D},\tau)=\vert \tilde{I}(f_\text{D},\tau) \vert^2$ is a function of the Fourier conjugates of time and frequency, which are called Doppler rate $f_\text{D}$ and delay $\tau$ respectively. These names are due to their geometric interpretation as differential Doppler shift and temporal delay of a signal. For a given lensing structure at angular coordinate $\bm{\theta}$ -- called image henceforth -- from the line of sight, these coordinates obey a parabolic relation to each other. This is the reason for the appearance of scintillation arcs. One way to parameterize these relations is
\begin{align}
    f_\text{D} &= - \frac{\nu}{c}\bm{V}_{\text{eff}} \cdot \bm{\theta} \, , \label{Eq:Def_fD} \\
    \tau &= \frac{1}{2c}D_\text{eff} \theta^2 \, . \label{Eq:Def_tau}
\end{align}
Here, $c$ is the speed of light, $D_\text{eff}$ is the effective distance and $\bm{V}_{\text{eff}}$ is the effective velocity. The effective picture of these quantities is an observer moving with $\bm{V}_{\text{eff}}$ and rays originating on a plane with distance $D_\text{eff}$ \citep[see][Figure 3]{2021MNRAS.500.1114S}. They can be translated to the distances $d$ and velocities $\bm{V}$ of the Earth ($_\oplus$), the scattering screen ($_\text{scr}$), and the pulsar ($_\text{psr}$):
\begin{align}
    D_\text{eff} &\equiv \frac{d_\text{scr} d_\text{psr}}{d_\text{psr}-d_\text{scr}} \, , \label{Eq:Def_Deff} \\
    \bm{V}_\text{eff} &\equiv \bm{V}_\oplus - \frac{d_\text{psr}}{d_\text{psr}-d_\text{scr}}\bm{V}_\text{scr} + \frac{d_\text{scr}}{d_\text{psr}-d_\text{scr}}\bm{V}_\text{psr} \, . \label{Eq:Def_Veff}
\end{align}
Because of the large distances, only velocity components perpendicular to the line of sight are relevant. Hence, all velocities are two-dimensional (2D) vectors.

Techniques not used in this work include direct measurement of the impulse response function using giant pulses \citep{2017ApJ...840L..15M} or cyclic spectroscopy \citep{2011MNRAS.416.2821D,2013ApJ...779...99W,2021ApJ...913...98D}. Instead, we will focus on secondary spectra, which map out the power belonging to structures on the screen \citep[][\cref{Eq:Def_fD,Eq:Def_tau}]{2004MNRAS.354...43W}.

\subsection{PSR B1508+55}

The pulsar B1508+55 (J1509+5531) is located in the constellation Draco and circumpolar at the Effelsberg telescope and the Low Frequency Array (LOFAR). It was discovered by \citet{1968Natur.219..576H} with a spin period of 0.739681922904(4) seconds \citep{2004MNRAS.353.1311H} and \citet{2015ApJ...808..156S} measured a dispersion measure (DM) of 19.6191(3)\,pc\,cm$^{-3}$. No orbital companion is known.

\citet{2005ApJ...630L..61C} found that B1508+55's perpendicular proper motion is one of the largest known with the most recent measured value $963^{+61}_{-64}$\,km\,s$^{-1}$ by \citet{2009ApJ...698..250C}. Origins of such high velocities were explored by \citet{2008MNRAS.385..929G}. \citet{2009ApJ...698..250C} determined this velocity by measuring the pulsar's distance $d_\text{psr}$ and its angular proper motions $\mu_\text{RA}$ and $\mu_\text{Dec}$ in right ascension and declination respectively:
\begin{align}
    d_\text{psr} &= 2.10^{+0.13}_{-0.14}\,\text{kpc} \, , \label{Eq:Chatterjee_dpsr} \\
    \mu_\text{RA} &= -73.64^{+0.05}_{-0.04}\,\text{mas yr}^{-1} \, , \label{Eq:Chatterjee_mua} \\
    \mu_\text{Dec} &= -62.65^{+0.09}_{-0.08}\,\text{mas yr}^{-1} \, . \label{Eq:Chatterjee_mud} 
\end{align}

Before the establishment of secondary spectra as the primary tool to study scintillation, the characteristic time and frequency scales of this pulsar's scintillation have been measured in a number of studies by \citet{1970MNRAS.150...67R}, \citet{1981MNRAS.194..623A}, \citet{1982Natur.298..825L}, \citet{1985MNRAS.214...97S}, \citet{1999ApJS..121..483B,1999ApJ...514..249B,1999ApJ...514..272B}, and \citet{2001Ap&SS.278...43K}. These studies showed that the scintillation scales of B1508+55 are unstable and strongly varying between observations. More recently, this result was summarized and confirmed by \citet{2018A&A...618A.186W,2018yCat..36180186W} who also published dynamic spectra at high frequencies of 2250\,MHz.

\citet{1992ApJ...392..530K} and \citet{2004A&A...425..949E} studied the long-term flux-density variation of the pulsar, where the former found an unusually long refractive time scale that was not present anymore in the data of the latter.

The first secondary spectra of this pulsar were then published by \citet{2007ASPC..365..254S}. They already show the same visual characteristics as the earliest observations done for this work. In particular, parallel stripes seem to be convolved with the scintillation arc. This feature does not belong to the expected phenomenology produced by a single thin screen and is thus the first major motivation for this study.

The second major motivation for this study are the results of \citet{2018evn..confE..17W} who studied echoes in the pulse profile at 150\,MHz observed with LOFAR. These echoes could be directly imaged via very long baseline interferometry (VLBI) revealing linearly aligned structures of one arcsecond in total length. The pulsar is moving along these structures and was expected to cross two brighter ones at some point in 2021 and 2023 respectively. \citet{2020ApJ...892...26B} independently observed the echoes below 100\,MHz using the LWA1 and also concluded scattering in the ISM as their origin.

In the most recent study of B1508+55's scintillation, \citet{2021MNRAS.506.5160M} observed at 550-750\,MHz simultaneously at the Giant Metrewave Radio Telescope (GMRT) and at the Algonquin Radio Observatory (ARO). Their secondary spectra also show the peculiar parallel stripes and they confirmed consistent movement of features between observations. Importantly, they found a low scintillation correlation between the sites as well as a travel time of the scintillation pattern that was much lower than the scintillation time scale and concluded the presence of a second screen in addition to the one causing the echoes.

\section{Observations and Data}
\label{Sec:data}

Our observations are from the Effelsberg 100-m telescope. While the main observing campaign started in March 2020 using the L-band receiver \textit{P217mm 7-Beam} (18-21 cm prime focus receiver), the settings were chosen under the impression of five early observations in November 2019 and January 2020 of which some used a different receiver -- the ultra broadband receiver (UBB) -- that also covered higher frequencies of up to 2600\,MHz. This study includes 76 distinct observations of 247 hours in total. 

The settings used for all following observations and the reduction of baseband data were optimized such that both axes of the secondary spectrum covered approximately twice the average region of signal, asserting sensitivity to a wide range of curvatures and possible outliers of higher delay. The usable band was 1.27-1.45\,GHz. The data were folded with 128 phase bins, 12.5\,kHz channels, and 10 second subintegrations with DSPSR \citep{2011PASA...28....1V}.

The folded data were then read to a 4D data cube of phase, polarization, frequency and time using PSRCHIVE \citep{2012AR&T....9..237V}, of which only the total intensity was used to continue. After dedispersing, the bandpass is determined as the median over time of the phase-averaged intensity and corrected for by division of the data by the bandpass.

The following recipe for noise removal was used: The off-pulse region was selected as the phase bins of the averaged pulse profile whose value is below the median. The standard deviation of that off-pulse region as a function of time and frequency is then used as a measure of radio-frequency interference (RFI). If this value is more than 10$\sigma$ above the mean corrected for outliers, it was masked. After that, the remaining noise was estimated by the mean over the off-pulse region for each bin of time, frequency, and polarization and subtracted accordingly.

Dynamic spectra were then created by multiplying along the phase axis by a new mean pulse profile created from the corrected data and subsequently summing over that axis. Since the overall intensity is irrelevant for studies of scintillation on the time scales regarded here, the dynamic spectra were finally divided by their standard deviation, which creates typical intensity values in the range of single-figure numbers. 

As can be seen in \cref{Eq:Def_fD}, the Doppler rate of a fixed angle depends also on frequency, which leads to broadening of scintillation arcs for wider bandwidths. To avoid this, the Fourier transform over time is rescaled to a common reference frequency. This method was introduced as the NuT transform in \citet{2021MNRAS.500.1114S}. It has the additional advantage that all spectra could be scaled explicitly to a frequency of $1400$\,MHz, avoiding errors in curvature measurement due to varying relative signal strength of different frequencies. Furthermore, to compute secondary spectra, the mean of dynamic spectra was subtracted, which reduces the influence of power in the centre.

\section{Phenomenology}
\label{Sec:Phenomenology}

\begin{figure*}
 \includegraphics[width=\textwidth]{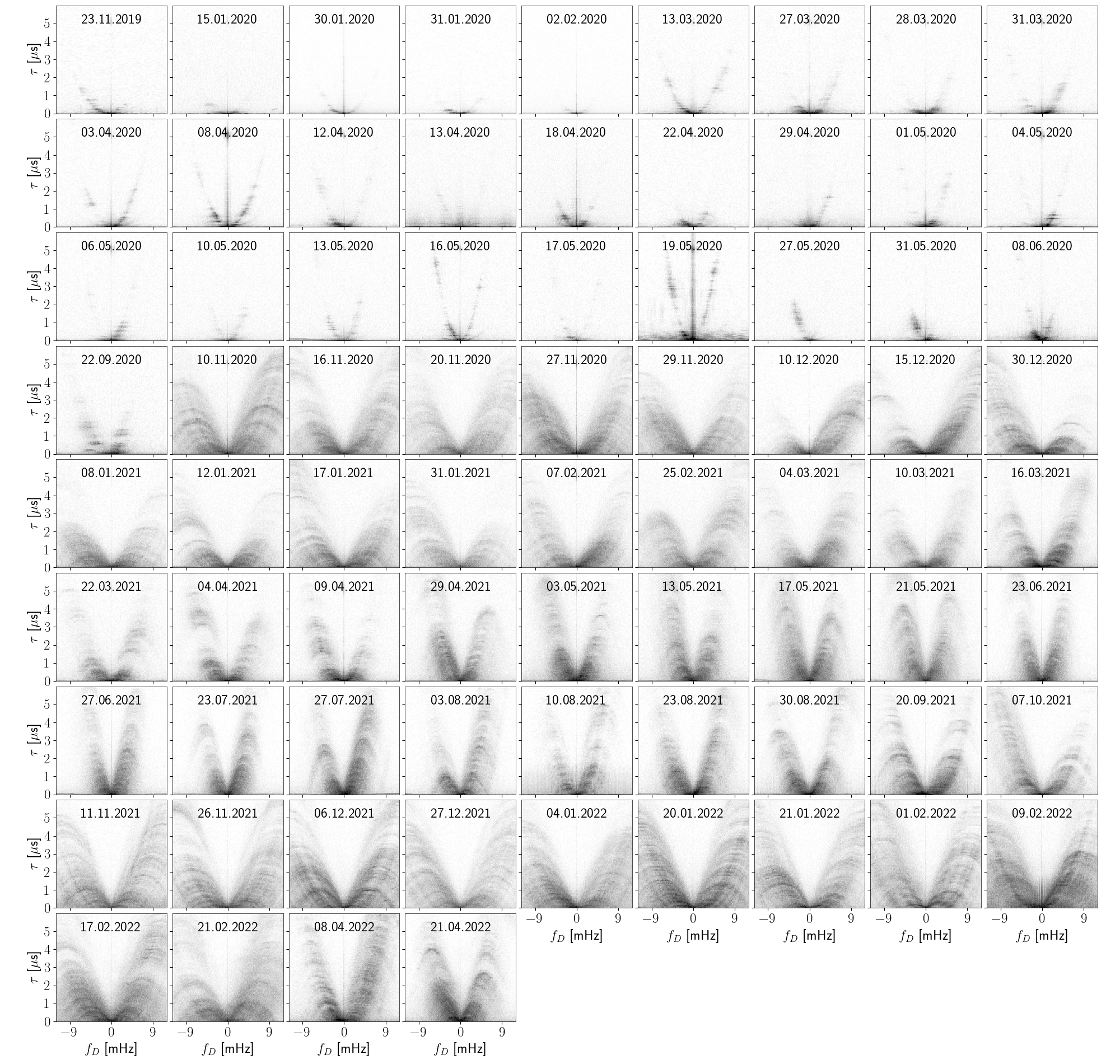}
 \vspace{-3mm}
 \caption{Secondary spectra obtained from observations of B1508+55 from the Effelsberg 100-m telescope for this study revealing a significant change in phenomenology in autumn of 2020. The dates are given in day, month, and year in this order. For this plot only, all secondary spectra have been rescaled to the same level of noise. In all plots of secondary spectra in this work, pixels below the resolution of the plot have been combined by taking the mean. The color scale is logarithmic and spans roughly four orders of magnitude. The spectra have been zoomed in -- the full axes for most data extend to $50\,\text{mHz}$ and $40\,\mu\text{s}$.}
 \label{Fig:RawData}
\end{figure*}

\begin{figure*}
 \includegraphics[width=\textwidth]{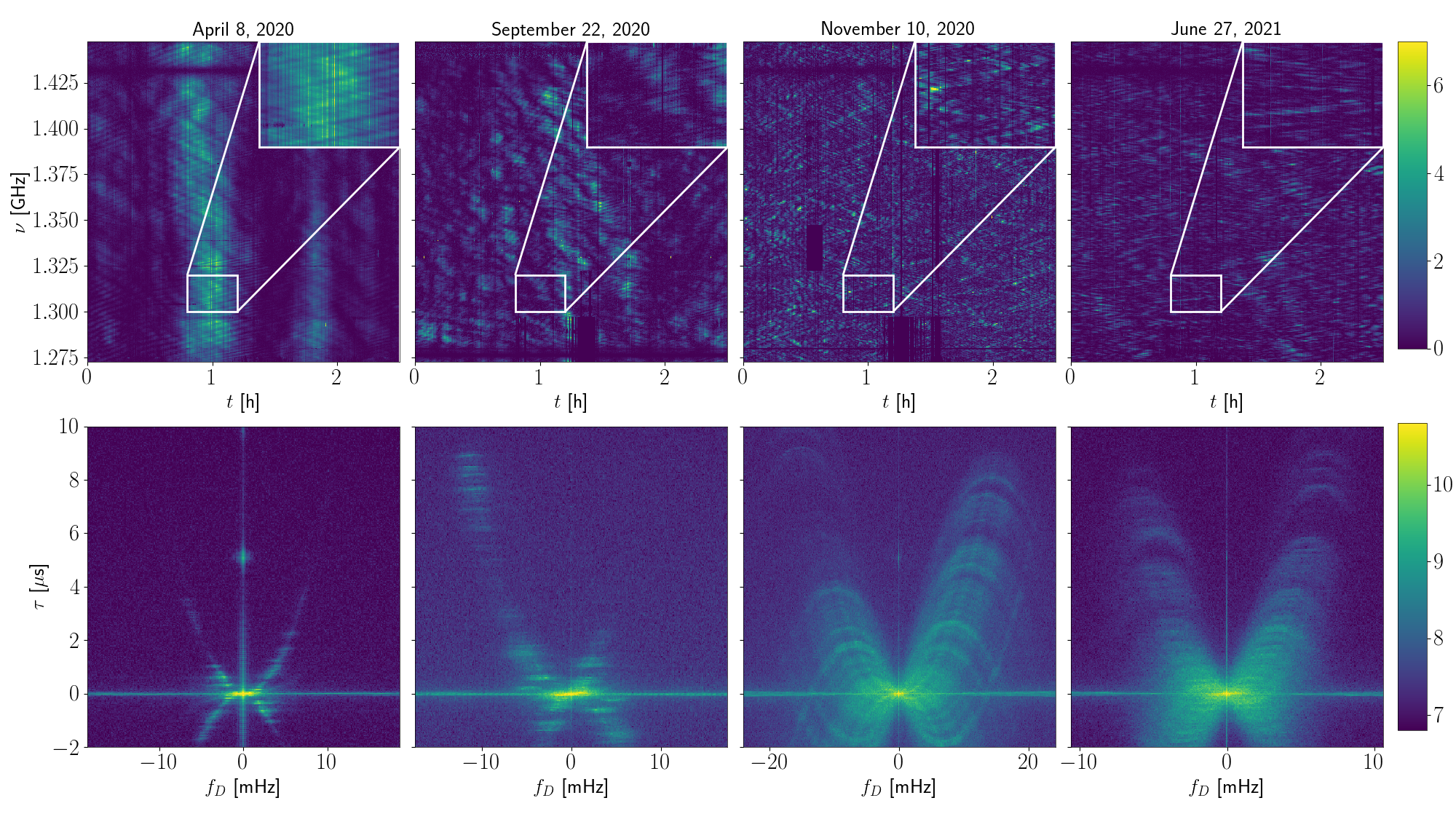}
 \vspace{-7mm}
 \caption{Examples of dynamic spectra (top row) and secondary spectra (bottom row) that have been chosen to highlight the different states of scintillation. The dynamic spectra have been cut only for the plots to produce comparable scales and the Doppler rate axes have been scaled such that the widths of the arcs are the same. The color scale is linear for the dynamic spectra and logarithmic in powers of ten for the secondary spectra.
 }
 \label{Fig:Transition}
\end{figure*}

\Cref{Fig:RawData} shows the secondary spectra of all observations from Effelsberg used for this study. To highlight the prominent phenomena and their evolution, characteristic examples of dynamic and secondary spectra are shown in \cref{Fig:Transition}.

\subsection{Terminology}

\subsubsection{Weak and Strong Scattering}

The terminology used here when referring to the strength of scattering needs to be distinguished from the strength of scintillation. Both terms have been used synonymously in the past \citep[e.g.][]{1990ARA&A..28..561R}. However, there are two measures of strength in the context of scintillation such that we chose to explicitely discriminate between them as follows. Weak scintillation manifests in intensity variations which do not modulate the intensity significantly compared to the mean intensity and results from weaker phase variations on the screen. Strong scintillation results in strong interference patterns that can completely extinguish the signal. PSR B1508+55 clearly exhibits strong scintillation. The further division into weak and strong scattering is essentially defined over the existence or non-existence of arclets in addition to the main arc within secondary spectra. Weak scattering is weak because the scattered images are much dimmer than the central direct line of sight, such that only interferences between this central image and the outer images are observable. In the case of strong scattering, interferences between all images are observable, without interference involving central images being orders of magnitude brighter.

\subsubsection{Arclets and Stripes}

As described above and first explained by \citet{2004MNRAS.354...43W}, arclets arise from the interference between images that are not at the centre of the distribution. Phenomenologically, they manifest as downward parabolae whose apexes align on the main arc which has the same curvature as the arclets. In this case, \cref{Eq:Def_fD,Eq:Def_tau} for the locations of power need to be updated to include the two angular coordinates in question:
\begin{align}
    f_\text{D} &= - \frac{\nu}{c}\bm{V}_{\text{eff}} \cdot \left(\bm{\theta}_1-\bm{\theta}_2\right) \, , \label{Eq:Def_fD_SS} \\
    \tau &= \frac{1}{2c}D_\text{eff} \left(\theta_1^2-\theta_2^2\right) \, . \label{Eq:Def_tau_SS}
\end{align}

The features we refer to as stripes are elongated parallel distributions of power in the secondary spectrum whose positions are aligned along a parabolic arc. Stripes were already observed for B1508+55 by \citet{2007ASPC..365..254S}. They are an unusual feature which has rarely been observed. \citet{2007A&AT...26..517S} found similar stripes for B0355+54 and \citet{2020MNRAS.496.5149S} observed B0834+06 showing a similar state of scintillation for a limited amount of time. They are interesting for the general understanding of pulsar scintillation because they are highly ordered structures that are not predicted naturally by the single thin screen theory without finetuning of two-dimensional structures that would not be expected to be stable over such a long time. On the other hand, they are regular and stable enough such that the mechanism behind them should involve few enough degrees of freedom to be accessible to measurements. 

Additional interesting aspects of the stripes are that their width is not observed to depend on the timespan of the dynamic spectrum as might be expected from a moving central bright image. Also, although the stripes of a single spectrum look very similar to each other, they cannot be simply deconvolved into a thin parabola and a single stripe. The stripes are thin and completely parallel in this observing band. Their brightness distribution can be shifted from the main arc but shows only minor scattering compared to the offsets of the other stripes. Their length has no clear cutoff and depends on the brightness of the individual feature and the data quality. However, beyond the overall brightness, their length does not depend on the location in the spectrum.

Although the appearance of stripes correlates with (almost) achromatic temporal variations in the dynamic spectra, these are unlikely to be a mere effect of pulse-to-pulse variations on larger time scales because they are missing in dynamic spectra of later observations where stripes and arclets coexist. Other reasons will be discussed below by showing that different scattered images of the pulsar exhibit different intensity variations, which contradicts an origin at the level of the pulsar's emission.

\subsection{Transition to new State of Scintillation}

The secondary spectra of B1508+55 observed until September 22 at Effelsberg differ strongly from those observed on November 10 and later at the same telescope (see \cref{Fig:Transition}).

Before the transition, the secondary spectra showed stripes. As there are no arclets, the initial state of B1508+55 can be described as a case of weak scattering.
Many of the observations even show a secondary arc. It becomes more visible the narrower the parabola of the primary arc. Rarely, hints of a third and fourth arc can be seen that are even wider than the first two ones.

Intriguingly, the first observation after the transition already shows clear and strong arclets which means the pulsar now experiences strong scattering. The stripes are completely gone at first. In later observations however, flat features reappear at the apexes of arclets. Nevertheless, the pulsar remains in a state of strong scattering. These additional stripes, which coincide with an overall blurring of all other structures, are only visible during times of a lower width of the arc. Due to the annual cycle of arc curvatures, this is happening during the summer months of each year (see \cref{Fig:RawData,Fig:Transition}).

\section{Scintillation Arc Evolution}
\label{Sec:Scint_Arcs}

The Earth's velocity vector changes its relative angle to the line of sight to the pulsar during the course of a year. Depending on the scattering geometry, this changes the time scale of scintillation (see \cref{Eq:Def_fD}) and hence alters the width of scintillation arcs. Measuring scintillation arcs in multiple observations during different times of the year can thus be used to infer a geometric model of the scattering material. Variations can also be caused by orbital motions of the pulsar. In this case they provide constraints on orbital parameters.

Recently, studies like \citet{2020MNRAS.499.1468M}, \citet{2020ApJ...904..104R}, \citet{2022MNRAS.511.1104M}, and \citet{2022ApJ...927...99M} have shown the potential of studying the temporal variation of scintillation arcs.

\subsection{Modeling}

Following \cref{Eq:Def_fD,Eq:Def_tau}, a one-dimensional (1D) distribution of scattered paths results in a parabolic distribution of power $\tau = \eta f_\text{D}^2$. The curvature parameter of scintillation arcs is given by
\begin{align}
    \eta \equiv \frac{c D_\text{eff}}{2\nu^2 V_{\text{eff},\shortparallel}^2} \, . \label{Eq:Def_eta}
\end{align}
$V_{\text{eff},\shortparallel}$ is the effective velocity vector defined in \cref{Eq:Def_Veff} projected onto the 1D distribution of images, i.e.~the screen's axis. The distribution of power can be more complicated, e.g.~a filled parabolic envelope (isotropic screen), downward arclets (interference between outer images, see \cref{Eq:Def_fD_SS,Eq:Def_tau_SS}), and multiple parabolae (multiple screens).

The curvature $\eta$ has become the standard parameter to characterize scintillation arcs although it has some disadvantages. First, it is frequency dependent. To get a single number characterizing the scintillation, it needs to be scaled to some reference frequency, which is typically chosen to be 1400\,MHz. Furthermore, $\eta$ is suboptimal for fitting since the region covered by the corresponding parabola within the secondary spectrum does not shift linearly with this parameter but rather with $1/\sqrt{\eta}\propto \vert V_{\text{eff},\shortparallel} \vert$. In particular, $\eta$ diverges in the important region around zero effective velocity. Finally, the parameter $\eta$ has no recognizable physical meaning by itself, which would help characterize a measured scintillation arc by labeling it with this number.

For all these reasons, we use a different parameter than the curvature $\eta$ in this work:
\begin{align}
    \partial_t \sqrt{\tau} = \sqrt{\frac{1}{2c}} \frac{\vert V_{\text{eff},\shortparallel} \vert}{\sqrt{D_\text{eff}}} = \frac{1}{2\nu\sqrt{\eta}} \, . \label{Eq:def_zeta}
\end{align}
The physical meaning described by the temporal derivative of the square root of delay -- which we refer to as the \textit{effective drift rate} -- is explained in \cref{Sec:Feature_Alignment}. The parabolic equation now becomes
\begin{align}
    \tau = \left( \frac{1}{2\nu}\times\frac{f_\text{D}}{\partial_t \sqrt{\tau}} \right)^2 \, .
\end{align}
This notation naturally leads to a linear dependence in $f_\text{D}$-$\sqrt{\tau}$ space which will be utilized throughout this work.

\subsection{Measurement}

\subsubsection{Weak Scattering with Stripes}

\begin{figure}
 \includegraphics[width=\columnwidth]{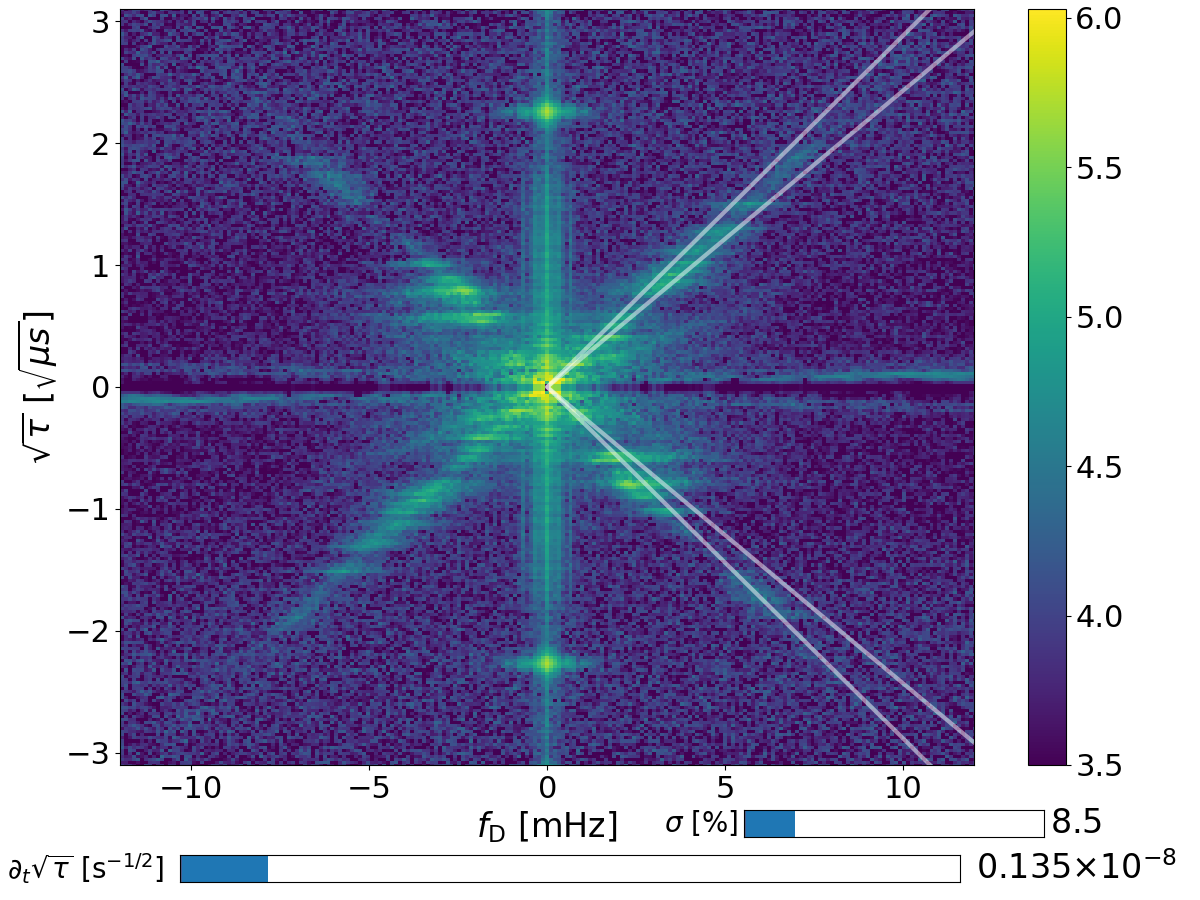}
 \vspace{-3mm}
 \caption{Example of manual arc detection using sliders and in $f_\text{D}$-$\sqrt{\tau}$ space that conveniently transforms arcs to lines. The color scale is in powers of ten and the white lines indicate the estimated accuracy of the manual measurement. This observation from April 8 in 2020 also shows a clear example of the secondary arc.}
 \label{Fig:Arc_manual}
\end{figure}

The stripes complicate the arc measurement significantly because their power distribution varies between observations and is not necessarily centered on the main arc which creates an asymmetry between the left and right part of the arc. Methods that rely on summing the intensity along hypothetical parabolae like employed in \citet{2020ApJ...904..104R} or on the identification of brightest points like in \citet{2020MNRAS.499.1468M} are not optimal in this case.

To avoid systematic errors caused by the lack of proper modeling of the stripes, we chose to measure the arcs manually, which is feasible with regard to the number of observations. In order to do this, the square roots of secondary spectra are transformed from $f_\text{D}$-$\tau$ space to $f_\text{D}$-$\sqrt{\tau}$ space for two reasons: First, this transformation visibly increases the signal to noise for the important regions of higher delay. By taking the square root, this transformation also achieves a uniform distribution of signal per area on the screen when pixels of the data before transformation are summed accordingly into pixels after the transformation \citep[compare similar transformations in][]{2021MNRAS.500.1114S}. Second, parabolae get transformed to straight lines which are more efficiently identified by the human brain. The same method was applied to the secondary arc in the cases it was sufficiently visible.

\Cref{Fig:Arc_manual} shows an example for such a manual measurement. The delay axis is resampled by summing over all pixels in $\tau$ that fall into a new pixel of $\sqrt{\tau}$ while correctly accounting for fractional coverage. The color scale is logarithmic. The linear relation searched for becomes
\begin{align}
    \partial_t \sqrt{\tau} = \frac{1}{2\times 1400\,\text{MHz}}\left\vert \frac{f_\text{D}}{\sqrt{\tau}} \right\vert \, .
\end{align}

Since the uncertainty of this method is dominated by unmodelled systematic deviations, a formal measure would underestimate the uncertainty. Instead, we chose the range in which the arc could reasonably be located by eye and used this as the standard error to be conservative.

\subsubsection{Strong Scattering}
\label{subsec:Arcs__StrongScintillation}

\begin{figure}
 \includegraphics[width=\columnwidth]{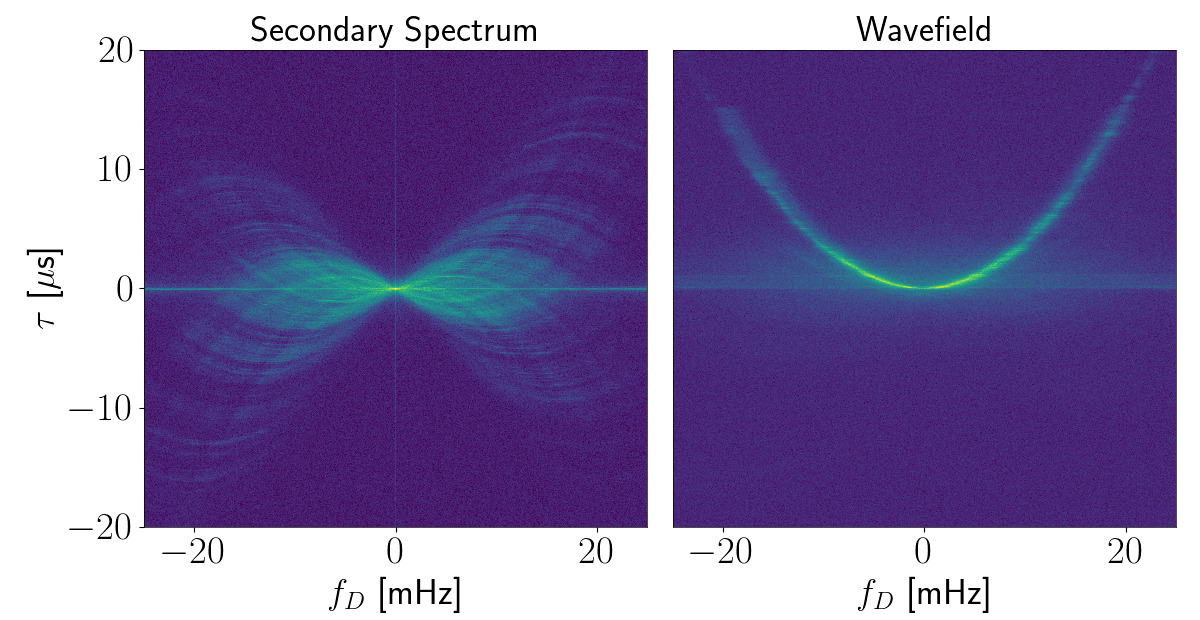}
 \vspace{-3mm}
 \caption{
Example of a wavefield obtained by phase retrieval as described in \citet{2022MNRAS.510.4573B}. On the left side the corresponding secondary spectrum is shown. This data was taken on December 6 in 2021. The color scale is logarithmic and different for both plots.}
 \label{Fig:Phase_Retrieval}
\end{figure}

While the presence of arclets introduces similar problems as the stripes for algorithms based on the distribution of intensity, it enables the option of using the eigenvector decomposition method introduced by \citet{2022MNRAS.510.4573B} in $\theta$-$\theta$ space \citep{2021MNRAS.500.1114S} which in principal can lead to a much higher precision than previously mentioned methods. 

They showed how it is possible to utilize also the phases of the conjugate spectrum that are usually dismissed by using its square -- the secondary spectrum. As a result, sufficiently stable eigenvectors can be found even from small subsets of the dynamic spectrum, such that arc curvature estimation and phase retrieval is possible. We applied this technique of phase retrieval to our data of B1508+55 and found convincing results for subset sizes in pixels of 
\begin{align}
   N_t = 60\times \frac{0.006 \sqrt{\mu s}/\text{h}}{\partial_t \sqrt{\tau}} ~~ , ~~ N_\nu = 200\, .
\end{align}
The change of scintle width was taken into account by correcting according to the manually measured scintillation arc width. Other parameters were the range of delays $\tau < 15\,\mu$s up to which corresponding angles $\theta$ are regarded for the eigenvectors and the number $N_\theta = 151$ of discrete steps over which to evaluate the eigenvector. These numbers where the minimum found such that artefacts and off-arc power were minimized in the wavefield that was computed from the square root of the dynamic spectrum with the retrieved phases applied. An example for an obtained wavefield is shown in \cref{Fig:Phase_Retrieval}.

\begin{figure}
 \includegraphics[width=\columnwidth]{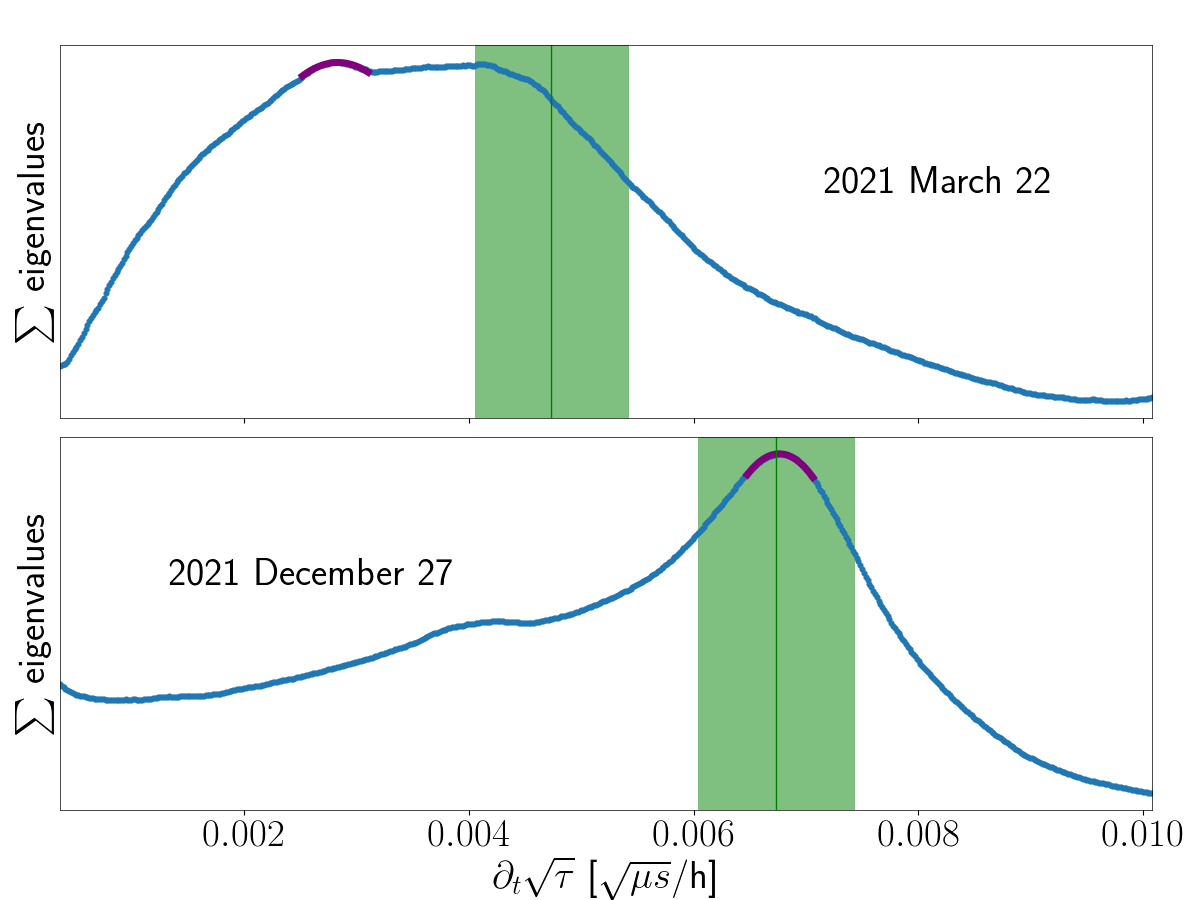}
 \vspace{-3mm}
 \caption{Examples of arc measurement using an adaption of the method of \citet{2022MNRAS.510.4573B}. The purple parabolas are fits to the peak of the sum of eigenvalues. The green region shows the manual measurement with estimated uncertainty. The two examples represent the extremes of failure and perfect match. The second peak becomes less prominent if the number of discrete steps on the eigenvector is increased but then gets broadened and further distorts the measurement.
 }
 \label{Fig:Arc_thth}
\end{figure}

However, we found that the method of finding the curvature belonging to the strongest possible eigenvector as described in \citet{2022MNRAS.510.4573B} deviated systematically from measurements done with the manual method described above. This deviation is strongest when the secondary spectra show reappearing stripes in addition to the arclets. We thus conclude that this method is also compromised by the peculiarity of this pulsar. Instead of resorting fully to manual measurements, we modified the eigenvector method to be more robust as described in the following. The original method relies on finding the largest eigenvalue over a distribution of curvatures evaluated on a small subset of data and then measuring the curvature by fitting those curvature measurements over frequency. In this case, individual eigenvalue distributions mostly did not follow a single peak distribution. Thus, we changed the order, sampled over the effective drift rate $\partial_t \sqrt{\tau}$ instead of the curvature $\eta$ to remove the frequency dependence and summed all eigenvalues before searching for the peak. Examples of the results are shown in \cref{Fig:Arc_thth}.

Still, this method remained biased because of the presence of a second peak that has no physical counterpart in the form of a secondary arc. This second peak is particularly prominent at lower numbers of $N_\theta$ than optimal for phase retrieval but also less wide such that its effect can be decreased slightly. Still, the main peak usually lies within the $1\sigma$ range of the manually determined result. Hence, the mean of both results was taken and the standard deviation of both measurements taken as the uncertainty of the combined measurement. Whenever the eigenvalue peak was not compatible with the manual measurement only the latter with its error was used. In cases where the deviation of both methods was smaller than the uncertainty of a parabola fitted to the peak, the uncertainty obtained from the latter was used.

\subsubsection{Additional Public Data}

Two recent publications contained scintillation arc data close to the time of the Effelsberg observations of this study and at close enough frequency bands such that they could be added to the analysis.

\citet{2021MNRAS.506.5160M} report data taken from the GMRT over 550–750\,MHz on August 6 and August 12 of 2017. They measured an arc curvature of $\eta\approx 0.76\,\text{s}^3$ for both observations. For both dates, we added this value to our analysis after applying the correct frequency scaling and estimating the uncertainty as the mean uncertainty of the Effelsberg weak scattering arc measurements since the data looked qualitatively similar.

\citet{2018A&A...618A.186W} report data taken from the Jiamusi 66 m telescope over 2180-2320\,MHz on December 13 of 2015 and October 31 of 2017. The corresponding dynamic spectra are publicly available. We took them from \citet{2018yCat..36180186W}\footnote{The corresponding link to the public data repository can be found in the bibliography.} and the secondary spectra we formed from these dynamic spectra revealed measurable scintillation arcs. The latter observation even shows a secondary arc as bright as the primary arc. We applied the same arc measurement technique as was used for the weak scattering Effelsberg data.

\section{Feature Alignment}
\label{Sec:Feature_Alignment}

The features of a scintillation arc that manifest in distinct blobs, arclets, or stripes can be translated to individual images of the pulsar. If the same features appear in several subsequent observations, they can be aligned, which provides a measurement of the movements of these features along the arc with respect to its origin.

\citet{2005ApJ...619L.171H} showed that the angles corresponding to different individual arclets o B0834+06 change by the same angular velocity between observations. They concluded that the scattering structures in the screen are approximately constant. \citet{2018MNRAS.478..983S} computed the angular movement of images for a specific model of the screen and predicted non-linear movement and even truncation of the images around the origin of the scintillation arc.

\subsection{Modeling}

\begin{figure}
 \includegraphics[width=\columnwidth]{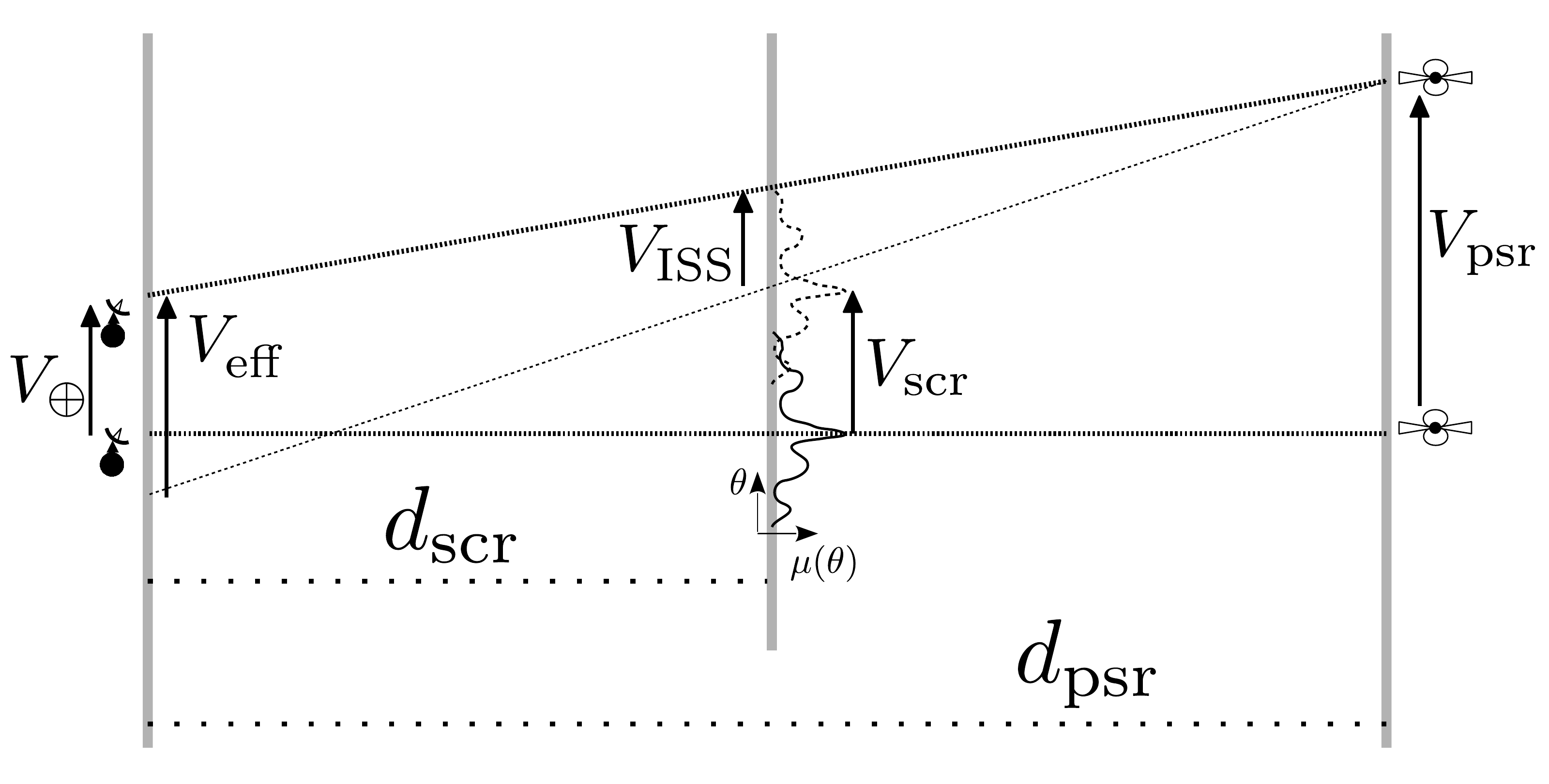}
 \vspace{-3mm}
 \caption{
The combined movements of Earth, ISM and pulsar result in a movement of the crossing point of the line of sight relative to the structures within the scattering screen. The finely dashed lines show the direct lines of sight from Earth to the pulsar at different times, while the coarsely dashed line shows the path crossing the same point in the image distribution on the screen after the movement.}
 \label{Fig:VelocityDerivation}
\end{figure}

The time derivative of the crossing point $\bm{\theta}_0$ with respect to some fixed location within the screen can be derived from geometrical arguments (see \cref{Fig:VelocityDerivation}). As an angular motion it is the physical velocity $\bm{V}_\text{ISS}$ of the scattering structure relative to the line of sight divided by the distance to the screen:
\begin{align}
    \frac{\der}{\der t} \bm{\theta}_0 = \frac{\bm{V}_\text{ISS}}{d_\text{scr}} \, . \label{Eq:los_movement}
\end{align}
Invoking the intercept theorem, this velocity can be derived:
\begin{align}
    \bm{V}_\text{ISS} = \frac{d_\text{psr}-d_\text{scr}}{d_\text{psr}}\bm{V}_\oplus - \bm{V}_\text{scr} + \frac{d_\text{scr}}{d_\text{psr}}\bm{V}_\text{psr} \, .
\end{align}
This term equals the relation used by \citet{2010ApJ...708..232B} and was branded \textit{scintillation speed} by \citet{1998ApJ...507..846C}.

Conveniently, the scintillation speed can be multiplied by $d_\text{psr}/(d_\text{psr}-d_\text{scr})$ to obtain the effective velocity and the remaining distance factors combine such that the result can be written in terms of effective quantities as defined in \cref{Eq:Def_Deff,Eq:Def_Veff} only:
\begin{align}
    \frac{\der}{\der t} \bm{\theta}_0 = \frac{\bm{V}_\text{eff}}{D_\text{eff}} \, . \label{eq:th0_dot}
\end{align}
Thus, the Doppler rate of a feature associated with an image at angular position $\bm{\theta}$ evolves in time as
\begin{align}
    f_\text{D} = -\frac{\nu}{c}\bm{V}_\text{eff}\cdot\left( \bm{\theta}-\int_{t_0}^{t}\frac{\bm{V}_\text{eff}}{D_\text{eff}}\der t' \right) \, ,
\end{align}
where $t_0$ is the point in time where $\bm{\theta}$ is defined as the current position of the image relative to the line of sight. However, if screens are indeed close to one-dimensional, images will never stay at a fixed position on the two-dimensional screen but rather move such that the line of images is never offset from the direct line of sight to the pulsar. This means that in practice only the angular coordinate relative to the screen matters and $\bm{V}_\text{eff}$ has to be replaced by its projection $V_{\text{eff},\shortparallel}$ onto the screen's axis.

This result can be shown to be identical to the formula
\begin{align}
    \frac{\der f_\text{D}}{\der t} = \frac{1}{2\eta \nu}\left( 1-\nu f_\text{D} \frac{\der\eta}{\der t} \right)
\end{align}
that has been used by \citet{2020MNRAS.499.1468M,2021MNRAS.506.5160M}.

As mentioned earlier, the effective drift rate $\partial_t \sqrt{\tau}$ is used in this study. The reason for using the temporal derivative of $\sqrt{\tau}$ instead of the temporal derivative of $f_\text{D}$ is to avoid a frequency dependence. Both quantities are linearly proportional to the angular position $\theta$ on the screen. Thus, the effective drift rate can be derived from \cref{Eq:Def_tau,Eq:los_movement}:
\begin{align}
    \tau &= \frac{D_\text{eff}}{2c}\left( \bm{\theta}-\int_{t_0}^{t}\frac{\bm{V}_\text{eff}}{D_\text{eff}}\der t' \right)^2 \\
    \Rightarrow ~~~ \partial_t \sqrt{\tau} &= \sqrt{\frac{1}{2c}} \frac{\vert V_{\text{eff},\shortparallel} \vert}{\sqrt{D_\text{eff}}} \, .
\end{align}

\subsection{Measurement}

\subsubsection{Weak Scattering}
\label{Sec:FeatureMovement_WS}

\begin{figure}
 \includegraphics[width=\columnwidth]{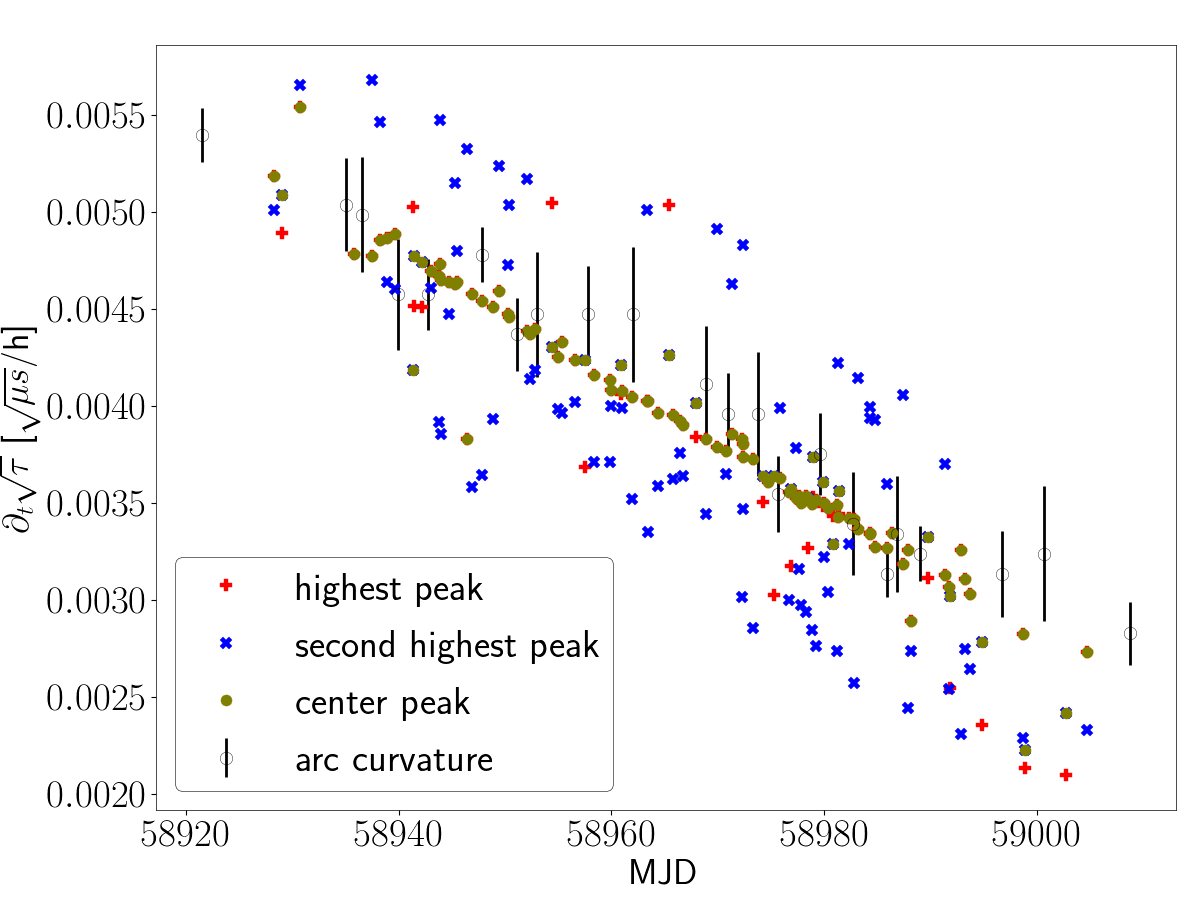}
 \vspace{-3mm}
 \caption{
Results of feature alignment using the peaks of template correlation. From the two highest peaks within a fixed range, the one closer to the mean of the results from arc curvature measurement of the two compared observations was chosen. This method is described in detail in \cref{Sec:FeatureMovement_WS}.}
 \label{Fig:Alignment_correlation_raw}
\end{figure}

\begin{figure*}
 \includegraphics[width=\textwidth]{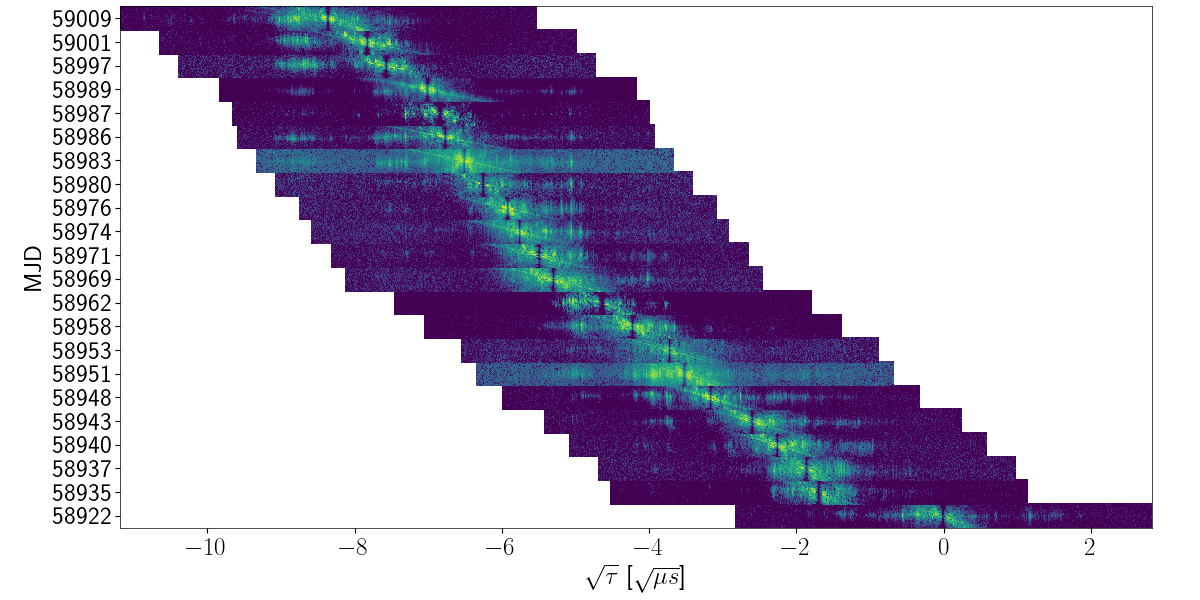}
 \includegraphics[width=\textwidth]{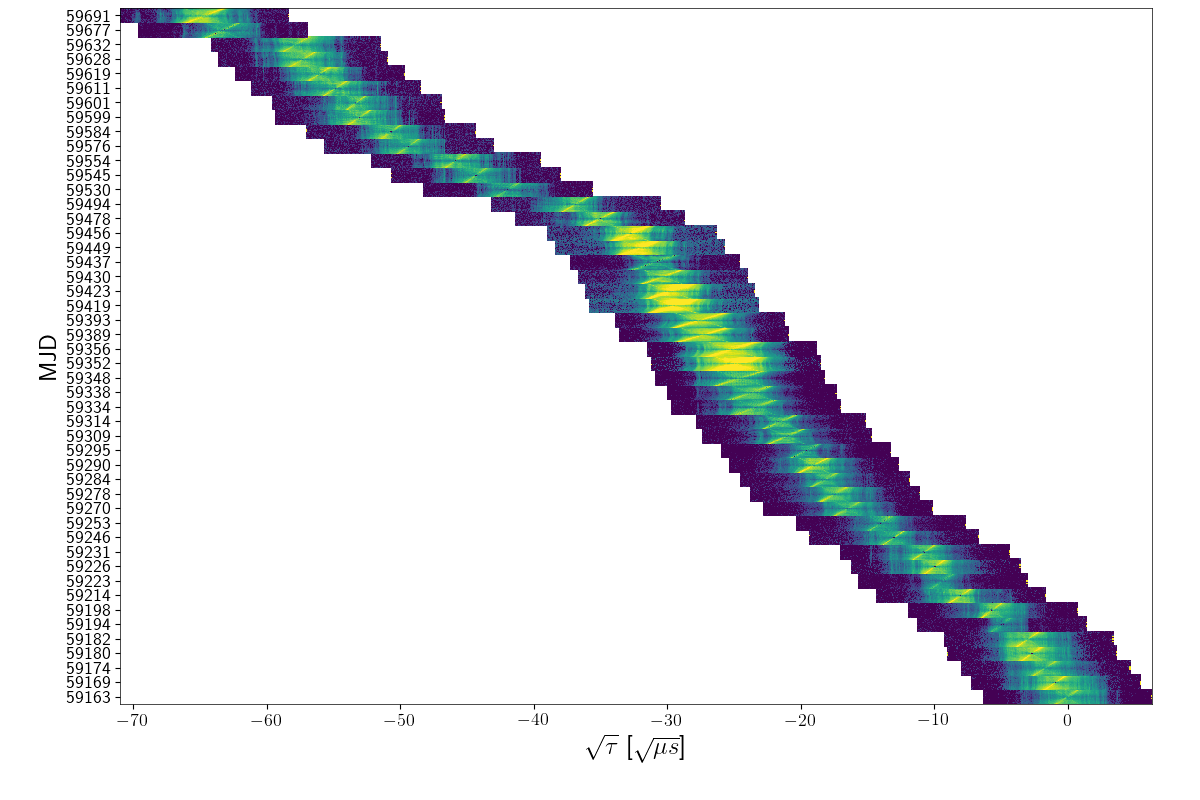}
 \vspace{-5mm}
 \caption{
Aligned spectra according to matching features. Shown are the coordinates along the main arc against the modified Julian dates of the observations.  Top: During the weak scattering regime it sufficed to only transform one axis, such that the unlabelled coordinate along the y-axis is $f_\text{D}$. Bottom: In the strong scattering regime the data had to be transformed to $\sqrt{\tau}$-$\sqrt{\tau}$ space (model-independent $\theta$-$\theta$) to transform arclets to lines. The contraction of the y-axes increases the remaining distortions due to reappearing stripes.}
 \label{Fig:Alignment_WS}
 \label{Fig:Alignment_SS}
\end{figure*}

In the case of weak scattering with stripes, aligning features across different observations is eased by the fact that the stripes are horizontal in delay and thus a feature can easily be identified with a single angular coordinate. To be model independent, angular coordinates should be expressed in $\sqrt{\tau}$. Since the features move along the arc, it is beneficial to shift the zero axis in Doppler rate to the centre of the arc at each delay. This step is only done to ease the comparability between observations. Although the centre of the arc depends on a measurement of that arc, this is not introducing a bias to the result of the alignment because the angular coordinate of a feature is not affected by this shift. The square root scaling of the delay can further be used to increase the signal to noise for high delay feature by summing the pixels such that a uniform pixel width is reached in $\sqrt{\tau}$-$f_\text{D}$ space is obtained.

To measure the effective drift rate $\partial_t \sqrt{\tau}$ between two observations, at least one common feature needs to be identified. In the case of weak scattering we employed a template correlation algorithm to find the best fitting shift. We chose a subset of the data with high enough cadence and then created $\sqrt{\tau}$-$f_\text{D}$ diagrams as described above. All diagrams are created in the same way:
\begin{enumerate}
    \item Take the amplitude of the conjugate spectrum, which equals the square root of the secondary spectrum.
    \item Compute coordinates in 800 equal steps in $\sqrt{\tau}$ ranging from $8$ to $0$\,$\mu$s as $-\sqrt{8}$ to $0\,\sqrt{\mu\text{s}}$ following the arc downwards for negative Doppler rates and from $0$ to $8$\,$\mu$s as $0$ to $\,\sqrt{8}\,\sqrt{\mu\text{s}}$ following the arc upwards for positive Doppler rates.
    \item Using the measured arc curvature, compute coordinates in $f_\text{D}$ ranging from $-2$\,mHz to $2$\,mHz in 40 equal steps after shifting the zero point of $f_\text{D}$ to $\sqrt{\tau}/\sqrt{\eta}$.
    \item Compute each pixel's value by taking it's range in $f_\text{D}$ and $\tau$ and summing over those values counting partially covered pixels by their fractional area covered.
    \item Only for the measurement of the the effective drift rate $\partial_t \sqrt{\tau}$, divide each diagram by the sum of all diagrams to correct for common artifacts and the brightening of features once they approach the origin.
\end{enumerate}
The shift between two observations is measured as follows:
\begin{enumerate}
    \item In the first observation, take a region centered on the arc with 10 pixels width in $f_\text{D}$ and 90 pixels width in $\sqrt{\tau}$ as a template to correlate with.
    \item The correlation with each region of the same size in the second region is computed. For each value of shift in $\sqrt{\tau}$ the shift with the highest correlation in $f_\text{D}$ is chosen to correct for slightly wrong arc curvature values.
    \item The above steps are repeated for the next template shifted by 10 pixels in $\sqrt{\tau}$. The new correlations are summed onto those already obtained.
    \item For the next step the search range of shifts is reduced to shifts corresponding to curvatures that do not deviate from the mean curvature of the two observations by more than 30\%.
    \item Within this region the highest two values are identified with the additional constraint of the second one being at least 5 steps apart from the highest one.
    \item If there is only one peak, it is chosen as the best fit value. If there are two peaks, the one closer to the mean curvature is chosen.
\end{enumerate}

The method of aligning features works for any pair of observations that is close enough in time to have detectable overlapping features. Here, we included every pair of observations that was separated by three weeks or less and obtain a point of data for each combination. \Cref{Fig:Alignment_correlation_raw} shows the results of this method compared to the allowed range and the values from the arc measurement. To use this data, its uncertainty has to be determined. This is not straightforward as the probability of two spectra being a match is distributed in a non-Gaussian way over all possible shifts. To estimate the uncertainty a posteriori instead, we subtracted a sinusoidal fit with a period of one year from the data and computed the uncertainty such that the variance is $1\sigma^2$ which corresponds to a normal distribution. The downside of this procedure is that the uncertainty estimate is model-dependent, which has to be kept in mind if a model is considered which is not linearly dependent on a velocity vector. This concern does not apply to this study though. Moreover, the data points are not completely independent because there cannot be more independent information than the number of observations because each of them resembles a fixed shift relative to a previous point in time.

\cref{Fig:Alignment_WS} shows the aligned $\sqrt{\tau}$-$f_\text{D}$ diagrams according to the measured shifts of adjacent observations. For the purpose of this plot the measured shifts from non-adjacent observations were ignored. In addition to the obtained effective drift rates, the alignment allows for two additional conclusions: First, the images fluctuate between observations, which means their amplitude is neither constant nor completely determined by their distance to the origin. Second, images can be identified over a long time, even after crossing the origin. This places constraints on plasma screen models because it requires quite different values of phase gradients to be present in close vicinity on the screen.

\subsubsection{Strong Scattering}

The parabolic arclets present in strong scattering pose a challenge for feature alignment because each feature is spread over a region in Doppler rate and delay that cannot be identified with a single angular position without using a measured curvature. Furthermore,in some observations stripes reappeared in addition to the arclets, which further distorted the position of features.

To make an alignment possible, the secondary spectra are transformed to incoherent $\theta$-$\theta$ diagrams as described in \citet{2021MNRAS.500.1114S}. Correlation peaks obtained by the technique used for the weak scattering data scattered randomly in the allowed range and thus allowed for no measurement. Therefore, we aligned the diagrams manually by plotting them next to each other and determining the best shift.

The obtained alignment is shown in \cref{Fig:Alignment_SS}. The uncertainty is even less constrained than in the weak scattering case because of the significant differences of the quality of alignments of pairs. Thus, we chose only to regard the results from the alignment of adjacent observations. The uncertainty depends on two factors: Firstly, there are noisy and perhaps also physical variations of the location of features relative to each other over time. This is estimated to be a constant number. Secondly, the uncertainty grows with diminishing overlap of observations. This is estimated as a linear dependence on the time between observations. The total error is the quadratic sum of these two errors. To fix the two free parameters, the constant error is estimated to be equal to the error obtained for the correlation method on the weak scattering data. The linear error is then fixed by again $\chi^2=1$ scattering around a sinusoidal fit.

\section{Eigenvector Decomposition Analysis}
\label{Sec:Eigenvectors}

\begin{figure}
 \includegraphics[width=\columnwidth]{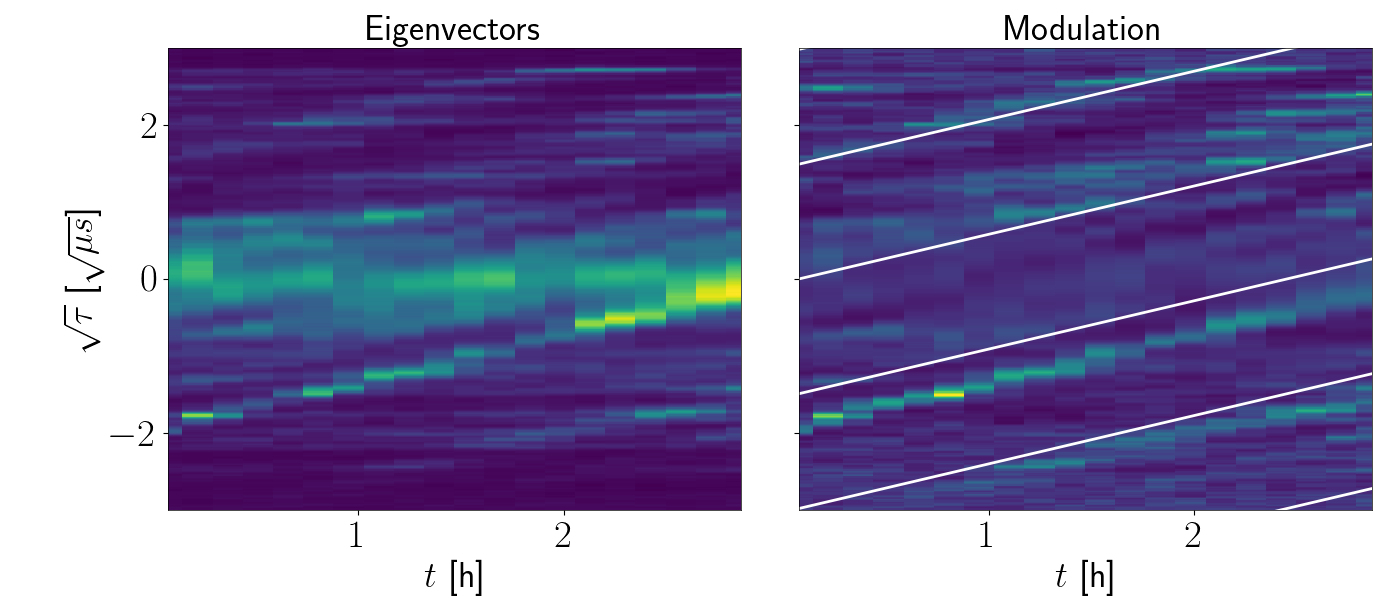}
 \vspace{-3mm}
 \caption{
Left: Amplitudes of eigenvectors over time and averaged over frequency. Although constant horizontal lines corresponding to the scattered images were expected, the data shows clear intensity modulations that travel along the images over time. Right: Same data divided by the time average of the eigenvectors. The white lines show the slope corresponding to the measured modulation speed. The data was taken on December 27 in 2021.}
 \label{Fig:Eigenvector_series}
\end{figure}

\begin{figure}
 \includegraphics[width=\columnwidth]{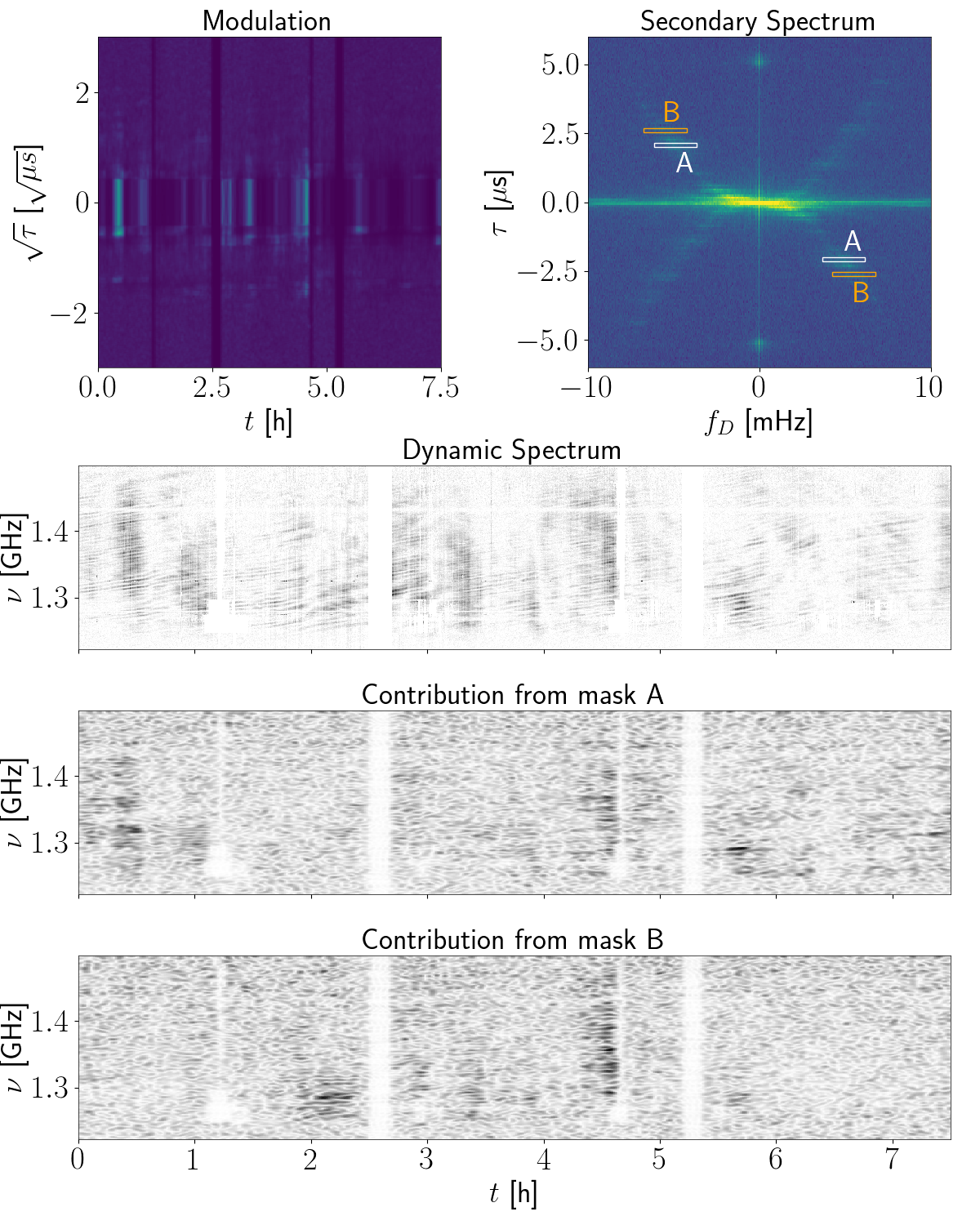}
 \vspace{-5mm}
 \caption{
Top left: Relative eigenvector evolution (see right panel in \cref{Fig:Eigenvector_series}) of data taken on April 12 in 2020 before the transition. Top right: Corresponding secondary spectrum. Below are shown in this order: The corresponding dynamic spectrum and the contributions to it obtained from inverting the Fourier transform after masking around the regions A and B in the conjugate spectrum. While the slope of modulation cannot be measured, a variation of the magnification of different images is clearly visible and shows that the almost achromatic structures in early dynamic spectra are caused by this modulation.}
 \label{Fig:backtrafo}
\end{figure}

Eigenvector decomposition is a powerful tool to obtain the brightness distribution of images on the screen. This technique is based on the $\theta$-$\theta$ transformation that was introduced by \citet{2021MNRAS.500.1114S} and transforms the secondary spectrum into a space where it, in the case of a single and one-dimensional screen, resembles a matrix equal to the outer product of the complex amplitude distribution $\mu(\theta)$ as a function of the angular coordinate along the screen's axis. It has been employed in \cref{subsec:Arcs__StrongScintillation} to measure the curvature of scintillation arcs.

This method can also be used to study the obtained brightness models for different parts of the data, as demonstrated by \citet{2021ApJ...907...49R}. In our case, the small required subset size allowed us to obtain a huge set of eigenvectors for every observation. They can best be visualized by taking the mean of their amplitudes over time or frequency, in order to suppress noise. An example of such is shown in \cref{Fig:Eigenvector_series}. In order to increase resolution, $\tau < 15\,\mu$s and $N_\theta = 601$ was used for this case. To assert statistical independence, the data subsets were chosen not to overlap here.

The eigenvector analysis reveals another special phenomenon of B1508+55: The brightness distribution is affected by a modulation that travels along the screen over time. The velocity of this modulation can be measured. To do this, we divided the frequency-averaged eigenvectors by their temporal mean. Then, the Hough transformation is applied to the eigenvectors over time to obtain a diagram over the $\theta$-intersect at $t=0$ and the slope of the modulation. Since not only lines of power but also lines of absent power exist, the most efficient way to find the best fitting slope is to take the peak of the standard deviation over the $\theta$-intersect as a function of slope. The result is quite stable for all observations after the transition, allowing only for annual variations of less than $0.05$\,$\sqrt{\mu s}$/h. If a constant value is assumed, the result for the \textit{modulation speed} of $0.6560(56)$\,$\sqrt{\mu s}$/h can be obtained by taking the mean and using the standard deviation divided by $\sqrt{N}$ as its uncertainty.

An alternative way of measuring this eigenvector modulation is to mask the arclet or stripe belonging to a single image, invert the Fourier transform, and average the amplitude of the result over frequency. This method yields the temporal evolution of eigenvectors without regard for the phases but slightly more stable in the case of weak scattering. Still, before the transition, the quality of eigenvectors is too poor to measure the modulation speed. Nevertheless, there are indications that the modulation is already present. An example is shown in \cref{Fig:backtrafo}. 

Since the modulation is traveling along the full visible screen within hours and is multiplicative and thus separable from the structure on the screen, its physical source is most likely located behind the screen causing the main arc. A modulation like this would naturally occur from a second screen if the extra delay is negligible such that there is no variation in frequency.

The vertical structures in the dynamic spectra of weak scattering could then be interpreted as the modulation of the bright central image of weak scattering. These structures correspond to the stripes in Fourier space. Thus, the stripes and the eigenvector modulation result from the same mechanism in this picture.

\section{Two-Screen Modeling}
\label{Sec:Two_Screens}

\citet{2006ChJAS...6b.233P} already reported on many pulsars with multiple scintillation arcs suggesting that scattering situations with more than one relevant screen are not unlikely. Two screens have been proposed as explanations for special scintillation phenomena -- for example by \citet{2016MNRAS.458.1289L}. 
Here, the evidence for two screens is very strong because such a theory consistently explains the special phenomena of stripes and eigenvector modulations (see \cref{Sec:Eigenvectors}).

Systems of multiple lenses have been studied theoretically by \citet{2022MNRAS.509.5872E,2020arXiv201003089F}. However, the stability of image positions we observed and the high number of images in the brightness distribution both imply that the lensing region of a single image is much smaller than the observable part of the screen. Thus, phenomena predicted by those studies on the scale of lenses that are resolved enough to be described analytically might not be accessible to our observations. On larger scales and without knowledge on the exact distribution of phases imposed by the screen, we can only rely on a number of data-driven approximations instead:
\begin{enumerate}
    \item \textbf{Anisotropy}: The phase screen is very anisotropic, i.e.~the phase varies exclusively in one direction on the screen.
    \item \textbf{Stability}: Along this axis of anisotropy, there are only certain points of stationary phase that allow propagation through the screen. On the large scales of interest, these points are stable in position over time and frequency.
\end{enumerate}
The second point can be interpreted as follows: An image exists in a region where the mean phase gradient is zero. On smaller scales within this region the distribution contains many slightly different phase gradients such that the combined emission of the whole region remains stable even if small changes in the geometry happen because of movements of observer and pulsar or slightly different frequencies used.

\begin{figure}
 \includegraphics[width=\columnwidth]{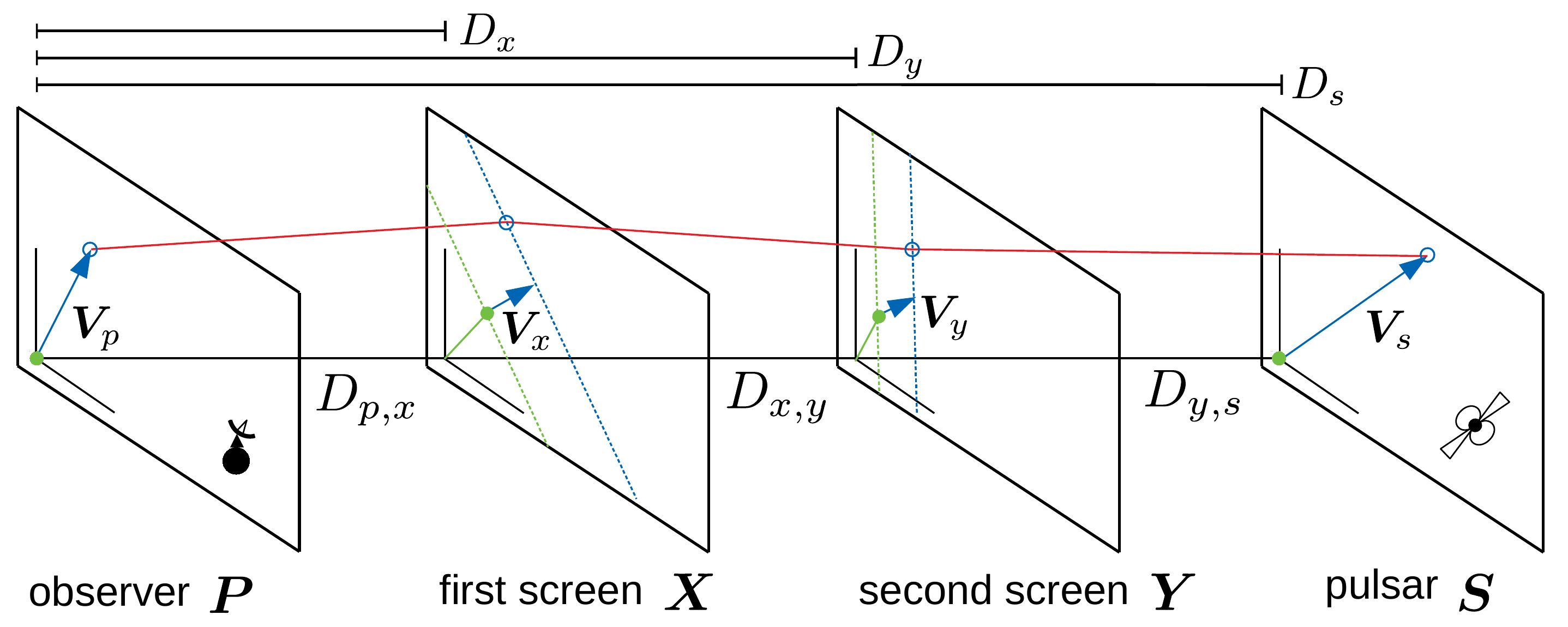}
 \vspace{-3mm}
 \caption{
The geometric situation of two scattering screens. The green dots show the points where the screens were intitially crossed while the blue circles show the current crossing point of the ray (red line) after some movement indicated by arrows. The green and yellow line represent a line of constant dispersive phase on the screens at initial and current time. In addition to the physical movement of the scattering material, the crossing point moves along these lines to the minimum of the total phase gradient.}
 \label{Fig:DoubleScreenVisualization}
\end{figure}

In \cref{Sec:Derivations}, we derive a double screen theory based on these approximations in order to model the phenomena we observed. \cref{Fig:DoubleScreenVisualization} shows the definition of distances and velocities in a two-screen geometry.

On each of the screens, propagation is limited to lines of scattering material with a suitable phase gradient. In the presence of a single screen, the dimension along these parallel filaments can be ignored because the radiation will always follow the direct line of sight in dimensions without variation in phase delay. In the case of two screens whose axes of anisotropy are tilted to each other, both dimensions have to be taken into account.

As a result, scintillation arcs and feature movements are the product of mutual interaction of the geometries of both screens. In addition, the screen closer to the Earth is illuminated not by a single point source but by multiple paths of propagation through the farther screen. Thus, a scintillation pattern is imprinted on the closer screen which can be observed by a variation of the amplitudes of images on the closer screen. If the variation in frequency can be neglected, the result is an eigenvector modulation like described in \cref{Sec:Eigenvectors}, whose modulation speed can be derived geometrically.

\section{Computational Methods}
\label{Sec:Computation}

\subsection{Modelling}
\label{Sec:Computation_Modelling}

We use coordinates in the equatorial coordinate system and define angles of velocities and screen orientations with respect to the axis of positive right ascension in mathematically positive direction on the sky. For each scattering screen $i$ in the model there are three free parameters: The distance $D_i$, the angle of anisotropy $\alpha$ and the local ISM velocity $V_{i,\shortparallel}$ along this axis of anisotropy. The direction of positive angles $\theta_i$ on the screen is arbitrary such that the same model is described when changing $\alpha_i$ by $\pm 180^\circ$ and inverting the sign of $V_i$.

We use \textit{Astropy} \citep{2018AJ....156..123A,2013A&A...558A..33A} to compute the velocity of the Earth projected on the sky towards PSR B1508+55 within this coordinate system. Accounting also for the additional velocity component from the rotation of the Effelsberg 100-m telescope around the Earth was found to be slightly below the measurement uncertainty and thus dismissed.

The secondary arc may relate to another screen than the one responsible for the eigenvector modulation. However, we will begin with the simplifying assumption that there are only two screens, and show that all of the observed phenomena in the data are fully consistent with this model.

The annual variation of the primary arc apparently changed consistently in phase at the point of the transition from weak to strong scattering (see \cref{Sec:Results}). There is no reason to assume that both screens changed properties simultaneously such that this shift of phase is also very likely originating from the first screen. The natural candidate for such a shift is the angle of anisotropy. A direct result of a shifted angle of anisotropy is a sensitivity to the second component of the velocity vector because the scintillation depends on the projection of the local ISM velocity onto the axis of anisotropy. Thus, the minimal model has eight free parameters: The distance $D_x$ of the first screen, the distance $D_y$ of the second screen, the angle $\alpha_{x,\text{WS}}$ of the first screen before the transition, its angle $\alpha_{x,\text{SS}}$ after the transition, the angle $\alpha_y$ of the second screen, the amplitude $V_x$ of the ISM velocity at the first screen, its angle $\alpha_{V_x}$, and the projected ISM velocity $V_{y,\shortparallel}$ at the second screen. Additional three parameters that will be marginalized over are the pulsar's distance and its velocity vector on the sky.

In addition, a one-screen model is explored for comparison. Again, only the angle of the screen is allowed to change with the transition. As this model does not give any prediction on the secondary arc or the modulation, these will be modelled by simple constants in order to compare the model to the full data. Thus, this model has seven free parameters.

In order to judge how well these models fare in general to describe the data, a simple heuristic model is introduced as a control. Here, three simple annual sine waves are used for the primary arc and its features before the transition, and separate ones for arc and features after the transition because there is a visible difference between those results. The secondary arc and the modulation are again modelled by constants. This model has eleven free parameters since each sine wave has three free parameters:
\begin{align}
    \partial_t \sqrt{\tau} = A \sin\left[ 2\pi(t-t_0)/(1 \text{year}) \right] + B \, .
\end{align}

\subsection{Parameter Inference}

\begin{table}
\begin{center}
\begin{tabular}{lcc} 
\toprule
data type & \multicolumn{2}{c}{$\partial_t \sqrt{\tau} ~ =$} \\
\toprule
 & One Screen & Two Screens \\
\midrule
primary arc & $\frac{\vert V_{\text{eff},\shortparallel} \vert}{\sqrt{2c D_\text{eff}}}$ & $\frac{\vert V_{\text{eff},x,\shortparallel} \vert}{\sqrt{2c D_{\text{eff},x}}}$ \\[8pt]
secondary arc & - & $\frac{\vert V_{\text{eff},y,\shortparallel} \vert}{\sqrt{2c D_{\text{eff},y}}}$ \\[8pt]
feature & $\frac{\vert V_{\text{eff},\shortparallel} \vert}{\sqrt{2c D_\text{eff}}}$ & $\sqrt{\frac{D_{\text{eff},x}}{2c}} \times \vert \dot{\theta}_{x,0} \vert$ \\[8pt]
modulation & - & $\sqrt{\frac{D_{\text{eff},x}}{2c}} \times \frac{V_m}{D_x}$ \\
\bottomrule 
\end{tabular}
\end{center}
\caption{Used formulae to model different data types using models of one or two screens. The relations for two screens are derived in \cref{Sec:Derivations}.}
\label{tab:fit_formulae}
\end{table}

We use the Markov Chain Monte Carlo (MCMC) ensemble sampler \textit{emcee} \citep{2013PASP..125..306F} for Bayesian inference of model parameters. The likelihood is constructed as independent Gaussian probability densities for each measurement. The modulation speed is lacking a reliable method of uncertainty estimation (see \cref{Sec:Eigenvectors}). Thus, a conservative assumption of using the standard deviation of all its measurements is used. This assumption is compatible with an approximately constant value, which is expected for a distant screen. In order to avoid double counting of information in the case of the pair alignments during weak scattering, the likelihood contributions of this data were weighted by the number of observations divided by the number of pairs. This is an approximate way to account for dependent measurements.

For the parameters $d_\text{psr}$, $\mu_\alpha$, and $\mu_\delta$ Gaussian priors are created according to the larger uncertainty of the two-sided uncertainties in \citet{2009ApJ...698..250C} (see \cref{Eq:Chatterjee_dpsr,Eq:Chatterjee_mua,Eq:Chatterjee_mud}). A Gaussian prior with a standard error of 100\,km/s is applied to all ISM velocities in order to keep values within a reasonable range of nearby star velocities as measured by \citet{2018A&A...616A..11G}. For all other parameters flat priors are used.

\Cref{tab:fit_formulae} summarizes the formulae used for different types of data for models of one or two screens.

\subsection{Simulations}

We perform simulations with the purpose of demonstrating that the theoretical framework described above can reproduce the observed phenomenology of B1508+55. Hence, dynamic spectra of same range and resolution as real observations are simulated. The parameters used equal the best fit values from the two screen model defined above.

The measurement of scintillation arcs, feature movement and modulation speed place no constraints on the distribution of images $\theta_i$ and their amplifications $\mu_i$. To ensure comparability between the observations, the same image positions were used for all simulations. Only on the first screen the amplifications of images were changed to replicate the change between weak and strong scattering. We found that the brightest central image could be kept in position and thus be treated in the same way as any other image to replicate the data.

Methods to infer information from image distributions are left for future work. Here only a proof of concept is desired. Thus, no attempt was made to exactly match the observed images.

Numerically, simulations were performed by computing the electric field as
\begin{align}
    E(t,\nu) = \sum_{m,n} \mu_m \mu_n \exp\left[ i\Phi(t,\nu,\theta_{x,m},\theta_{y,n}) \right]
\end{align}
on a grid of time and frequency. For the phase $\Phi$, \cref{Eq:Phase_abbr,Eq:2scr_phitilde} were used. The dynamic spectrum is obtained by taking the square modulus, after which the secondary spectrum can be obtained as usual.

Using the same methods as for the real data, the eigenvector modulation and its speed can be measured for the cases where the first screen is strongly scintillating. Here, the measured modulation speed can be directly compared to the theoretical expectation.

\section{Results}
\label{Sec:Results}

\subsection{Model Fits}

\begin{figure*}
\centering
 \includegraphics[width=\textwidth]{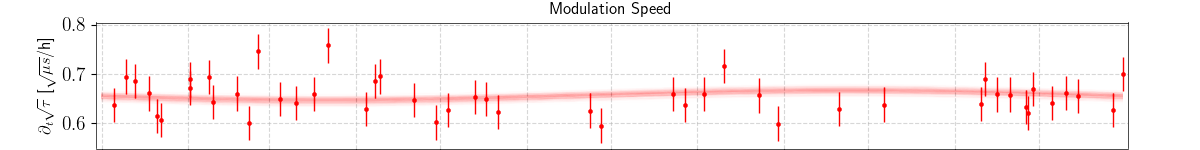}
 \includegraphics[width=\textwidth]{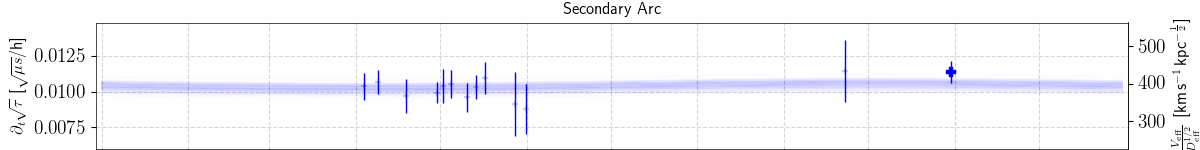}
 \includegraphics[trim={0cm 0.6cm 0cm 0cm},clip,width=\textwidth]{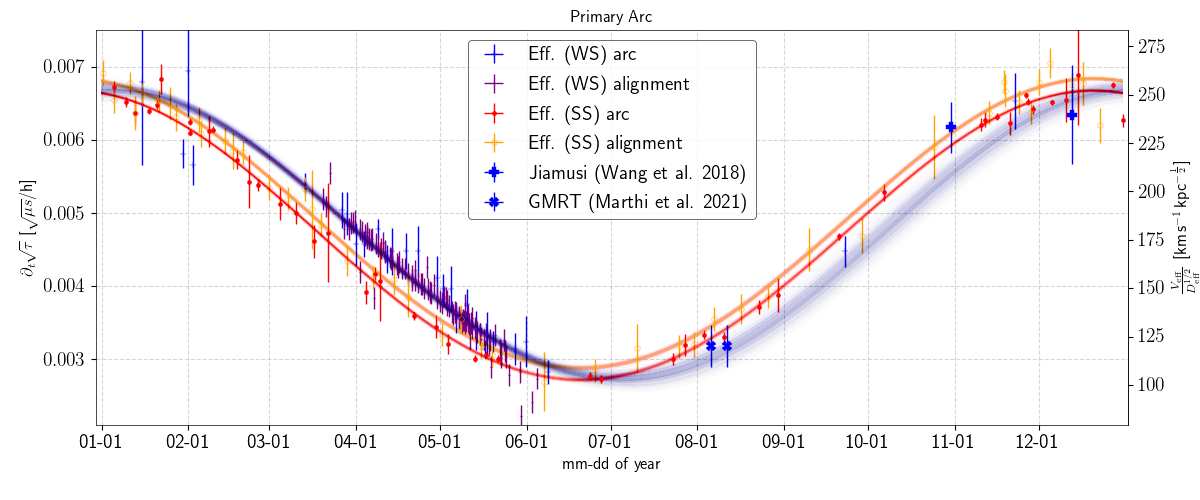}
 \vspace{-3mm}
 \caption{
 Measured values for the effective drift rate for both arcs and the modulation as well as models according to parameters drawn from the posterior. All times are shifted to the reference year 2020. (WS) and (SS) indicate data taken before the transition (weak scattering) and after the transition (strong scattering). 400 models are shown to illustrate the uncertainty whereas each line has the same color as the data it belongs to, except for the Jiamusi and GMRT data which belong to (WS) arc data. On the right axis, the effective drift rates are translated to the effective velocity over square root of effective distance.
 }
 \label{Fig:MCMC_results}
\end{figure*}

\Cref{Fig:MCMC_results} shows the accumulated results of all measurements of scintillation arcs, feature movements, and modulation speed. All observables can be expressed in effective drift rates. In addition, model predictions drawn from the posterior of the two-screen model are shown. The modulation speed and the secondary arc only show small annual variation. The primary arc before the transition clearly does not agree with the primary arc after the transition. While the feature alignment is compatible with the scintillation arc before the transition, there is a slight offset between them after the transition. The two-screen model is able to reproduce all of the measurements.

\begin{table*}
\begin{center}
\begin{tabular}{lccccccccc}
\toprule
model & $\alpha_x^\text{WS}$ [$\deg$] & $\alpha_x^\text{SS}$ [$\deg$] & $\alpha_{V_x}$ [$\deg$] & $D_x$ [pc] & $\vert V_x\vert$ [km/s] & $\alpha_y$ [$\deg$] & $D_y$ [pc] & $V_{y,\shortparallel}$ [km/s] & $\chi^2/\text{ndf}$ \\
\midrule
two-screen & 51.2(2.0) & 37.60(40) & -11(17) & 127.1(1.5) & 7.3(1.4) & -39.8(2.0) & 1940(123) & -02(55) & 0.95 \\
one-screen & 50.5(1.8) & 37.80(42) & 87(24) & 127.6(1.5) & 4.4(1.9) & - & - & - & 0.99 \\
heuristic & - & - & - & - & - & - & - & - & 0.93 \\
\bottomrule 
\end{tabular}
\end{center}
\caption{Best-fitting parameters and standard deviation of posteriors of the two-screen and one-screen model as well as a heuristic reference model. The last column lists the reduced $\chi^2$. These fit parameters are defined in \cref{Sec:Computation_Modelling}.}
\label{tab:result_table_2scr}
\end{table*}

\Cref{tab:result_table_2scr} shows the best-fitting values for the parameters of the two-screen and one-screen model with the standard deviation of the posterior distribution in parenthesis. The $\chi^2$ per number of degrees of freedom is also shown to compare the models. To provide a baseline in case of wrongly estimated uncertainties, the $\chi^2/\text{ndf}$ is also shown for the heuristic model. The pulsar's distance and velocity are not given because they could not be constrained any further than their priors.

The two-screen model is the only model that gives a physical prediction for all observables. At the same time it fits the data better than the one-screen model and the comparison to the heuristic model does not suggest the existence of badly modelled parts of the data. Even the small deviation between the effective drift rates of scintillation arc measurements and feature movement measurements during the regime of strong scattering is covered by the two-screen model. The only other explanation for this shift would be an undiscovered systematic error as such a deviation is impossible in a one-screen model.

These numbers can be used to infer physical scales from the measurements. According to the best-fitting two-screen model an arc containing power up to 10\,$\mu$s of delay corresponds to structures within a range of two astronomical units and the modulation speed translates to a velocity of about 8600\,km/s. The illumination pattern needs roughly 10 hours to travel along a screen like this.

If the data for the secondary arc is omitted in the likelihood, the best-fitting model still predicts values for the secondary arc that are compatible with the data, although less constrained. The same is true if the modulation is omitted in the likelihood. This finding provides further indication that the two phenomena originate from the same second screen, which can also be reproduced in simulations.

\subsection{Echo Crossing and Screen Orientation}

\begin{figure*}
 \includegraphics[trim={0.3cm 1cm 0.3cm 0cm},clip,width=\textwidth]{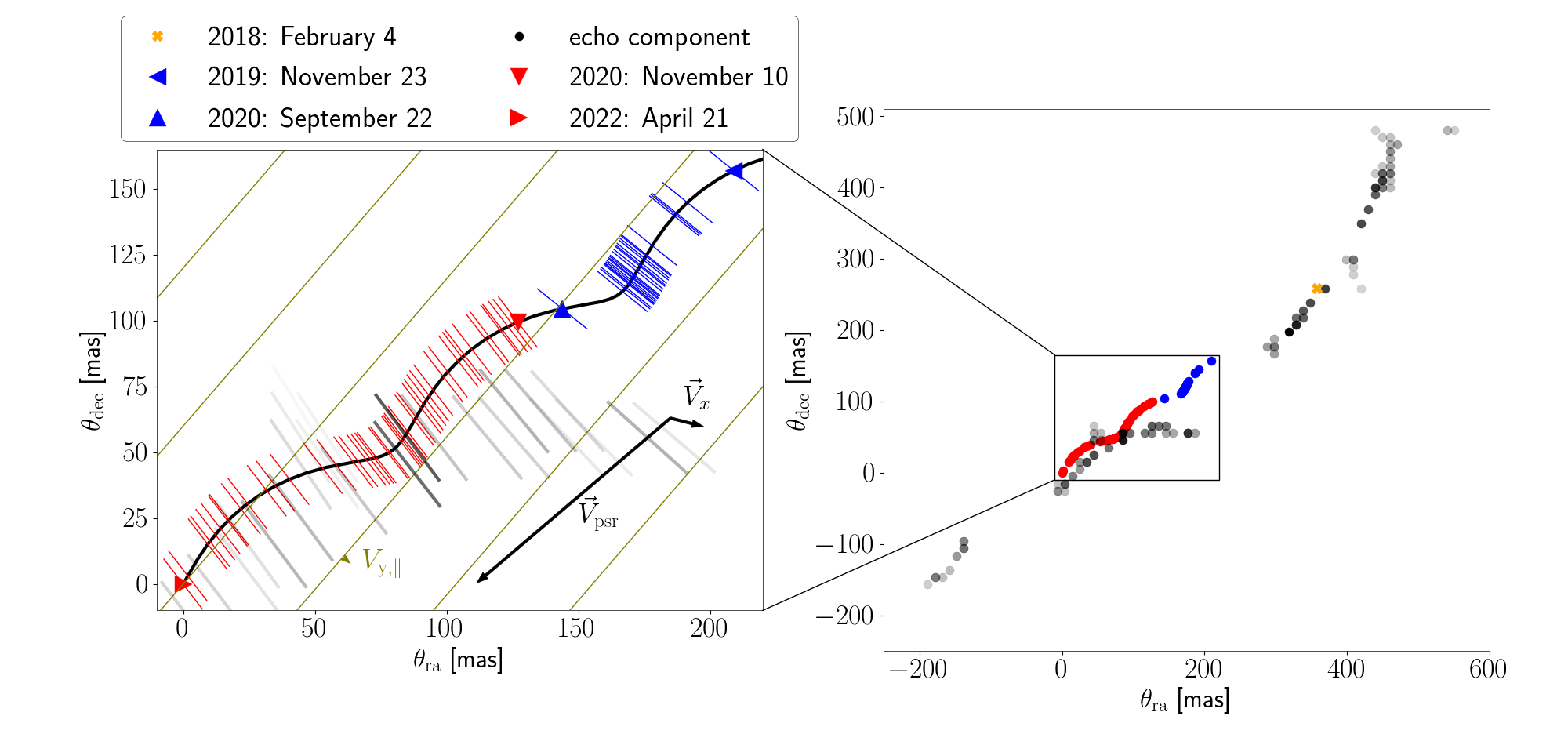}
 \vspace{-3mm}
 \caption{
The movement of the line of sight on the sky with the origin being centered on the last observation. The black line shows the points where the line of sight crossed the first screen in the past. The arrows' lengths correspond to the angular movement within one year. The lines placed at each observation show the direction of constant phase (perpendicular to the axis of anisotropy). Observations before and after the transition are shown in blue and red. The olive lines show the direction of constant phase of the second screen. The right plot shows the location of the echo components by \citet{2018evn..confE..17W}. The inferred direction of constant phase is also shown in the main plot. The echoes agree with the time of the transition and its change of angle.
}
 \label{Fig:los_movement}
\end{figure*}

A useful side effect of a change of the direction of anisotropy is sensitivity to the 2D vector of ISM velocity at the first screen. This means that the point where the direct line of sight to the pulsar crossed the screen in the past can be reconstructed unambiguously. The result is shown in \cref{Fig:los_movement}. After integrating the movement backwards to the time \citet{2018evn..confE..17W} observed B1508+55 with LOFAR, the localized echo components can be compared with the observations at Effelsberg. 

The result is an agreement with our model in three aspects. The transition coincides with the beginning of bright outer components. Furthermore, since an image forms where the phase is stationary, the orientation of constant scattering material has to be perpendicular to the line to the origin of the observation and can thus be computed. Right at the transition, this orientation changes gradually from the initial angle to the one after the transition. Finally, the distance of the screen as measured by \citet{2018evn..confE..17W} is compatible with our result for the first screen. Hence, the echoes can be explained by an effect of the same scattering screen on a larger scale.

Noteworthy properties of the best-fitting model are that the two screens are nearly perpendicular to each other while the second screen is close to parallel to the pulsar's velocity vector. This might be a reason for the images on the second screen being stable enough over time such that two-screen phenomena are so prominent in the scintillation of B1508+55. Short-lived images would produce an effectively comoving central image like described at the end of \cref{Sec:Two_Screens} which would reduce the first arc's observables to being compatible with a one-screen theory. However, the modulation of eigenvectors would still be visible. Such cases may be realized in many other pulsars or in the faint additional arcs of B1508+55.

\subsection{Comparison of Data and Simulations}

\begin{figure*}
 \includegraphics[trim={0cm 1.3cm 0cm 0.6cm},clip,width=\textwidth]{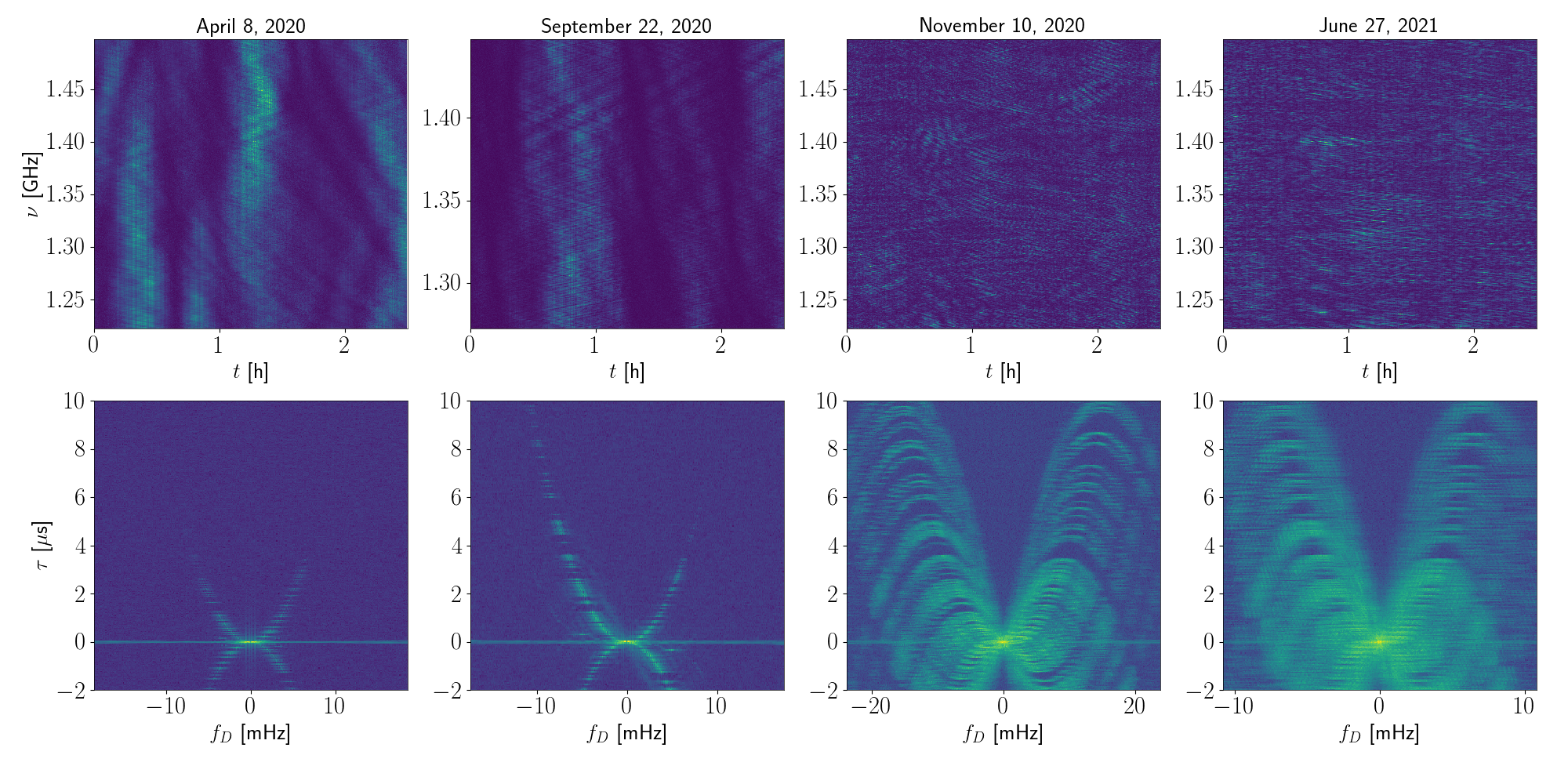}
 \vspace{-3mm}
 \caption{Examples of simulated dynamic spectra (top row) and corresponding secondary spectra (bottom row) using the model predictions at the corresponding dates. Noise and pulse-to-pulse variation have been added to make the results compatible to \cref{Fig:Transition}.
 }
 \label{Fig:Transition_sim}
\end{figure*}

\Cref{Fig:Transition_sim} shows simulations according to the best-fitting two-screen model. This is another important check if the two-screen model can potentially explain all the phenomena because the shift in appearences of stripes and arclets was not used for the measurements the model was fitted to. By changing the relative amplitudes of images on the first screen, it is indeed possible to reproduce stripes, secondary arcs and a shift from weak to strong scattering. 

Only for the observation on September 22 in 2020 one bright image had to be added to the second screen to reproduce the bright observed feature on the secondary arc. This is indicative of significant evolution of structures on the second screen, which is unfortunately impossible to quantify due to the secondary arc being fuzzy in all other observations. Even the reappearance of stripes on top of arclets during times of narrower scintillation arcs naturally occurs. This last effect does not even need a change of any of the image amplitudes.

One interesting observation is that to reproduce the last observation before the transition (September 22, 2020) it was necessary to make images on the incoming side much brighter than on the outgoing side. It is possible to interpret this as the incoming of a region of stronger scattering which later is responsible for the strong scattering and may even be identified with the onset of one of the echoes studied by \cite{2018evn..confE..17W,2020ApJ...892...26B}.

\begin{figure}
 \includegraphics[width=\columnwidth]{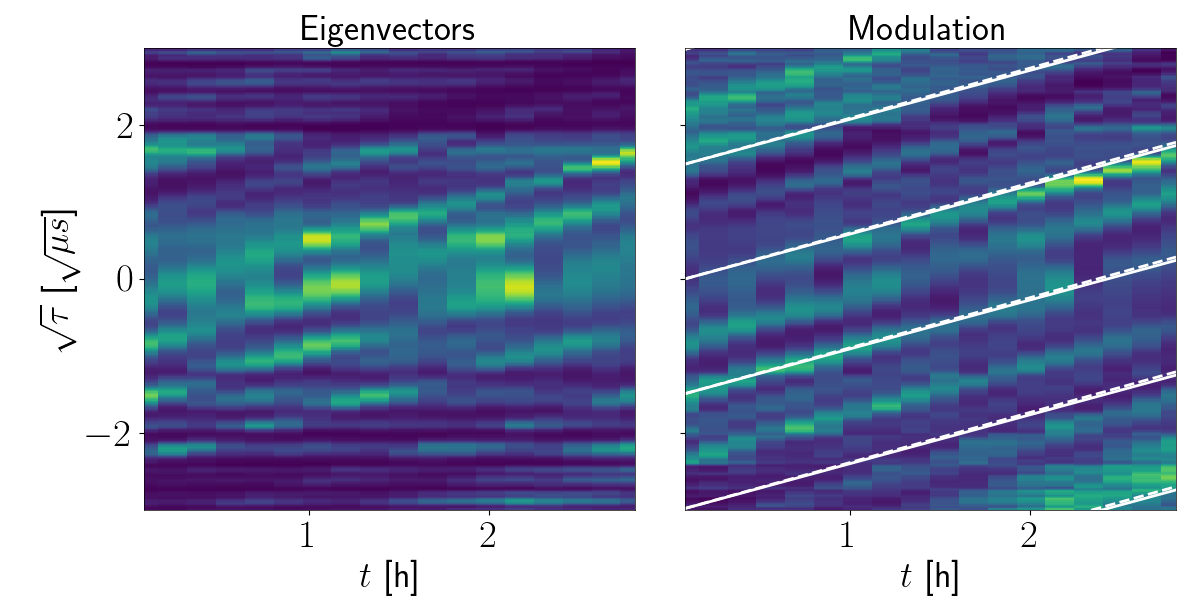}
 \vspace{-3mm}
 \caption{
Eigenvectors and their modulation inferred from the simulated data for November 10 in 2020. The white lines of measured and predicted modulation speed fall onto each other.}
 \label{Fig:simulations_modulation}
\end{figure}

As shown in \cref{Fig:simulations_modulation}, the simulations also reproduce the eigenvector modulation. Since the dynamic spectra were simulated, and not the eigenvector modulation directly, the method of measuring the modulation speed can be validated.

\section{Conclusions}
\label{Sec:Conclusions}

We reported a series of observations of PSR B1508+55 using the Effelsberg 100-m telescope. These observations covered roughly two and a half years and showed a large variation of scintillation arcs. Furthermore, in the secondary spectra we observed a transition from a weak scattering phase with multiple arcs and hitherto unexplained stripe-like features to a strong scattering phase with clear arclets.

The aim of this paper is to develop a model of the scattering geometry leading to the peculiar scintillation of the pulsar B1508+55, and thereby to introduce a number of new techniques for data analysis. We leave the physical interpretation of the scattering material as well as possible implications on properties of the pulsar itself for future studies. It shall be mentioned, though, that there are candidates: An A2 star is located 1.37\,pc from the line of sight to the pulsar at a distance of $120\pm8$\,pc \citep{2018evn..confE..17W}. Also, the edge of the local hot bubble lies at about 100-150\,pc away from us in the direction of the pulsar \citep{2017ApJ...834...33L}. Both are consistent with the distance of the closer screen. For the more distant screen, there is little known material along this line of sight that could be associated with it.

In addition to applying recently developed methods to measure scintillation arcs, we introduced methods to infer also independent information from the movement of features within the arc. The alignment of features over months showed consistent movement through the full arc, which is in disagreement with the model by \citet{2018MNRAS.478..983S} and indicates the presence of very small regions on the screen that contain a huge range of different dispersive phase gradients.

For the first time, we found a third observable by measuring the temporal evolution of the amplitude of images. This modulation speed could be measured for all observations that included arclets and provided a strong indication to interpret the scattering as a system of two screens.

To test this theory, we developed an analytic two-screen model. This model was then used to predict the observables as well as to simulate dynamic spectra. For comparison, we modelled all applicable phenomena also in the well-tested single-screen framework. Finally, we performed Bayesian inference on the obtained data and compared the models as well as the ramifications of the obtained parameters. We found that the best-fitting model agrees well with the VLBI results of echo components as observed by \citet{2018evn..confE..17W}. \citet{2021MNRAS.506.5160M} had previously suggested a two-screen model to explain the dynamic cross-spectrum seen on a 10000km baseline; our results strongly support this claim, and have allowed to measure the distance and orientation of both screens.

Although the two-screen model is able to reproduce the observations we reported on as well as observations by others, much of the underlying physics remains to be uncovered. In particular, the coinciding occurrence of the echo crossing, the change of the axis of anisotropy to being parallel to the pulsar's motion, and the onset of brighter scattered images suggest a common physical origin. Furthermore, the existence of strong two-screen effects in a system whose second screen is perpendicular to the first screen and aligned with the combined velocity of the pulsar and the first screen might be more than a coincidence.

Future study of this pulsar will include the analysis of data taken at LOFAR at a much lower frequency in order to study the validity of the model under these circumstances and to further explore the connection of the transition to the echoes that were discovered at these frequencies.

The transition underwent by B1508+55 allowed us to observe its scintillation through the lens of two different sets of phenomena which enabled a robust test of a two-screen scattering model. Since this explanation worked here, it is reasonable to assume similar models for other scintillating pulsars. Furthermore, we discovered the phenomenon of eigenvector modulation, which is a consequence of two-screen scintillation. In this case it was so strong that it also affected the shape of the secondary spectra by introducing the stripe feature, but it may reveal itself also in sources which otherwise appear dominated by one strongly scintillating screen. We have shown how to combine measurements of scintillation arc curvatures, feature movement along these arcs, and the speed of modulation along the features on these arcs and that the latter two can add a lot of information to the well-tested study of scintillation arc curvatures.

Scattering screens seem to be ubiquitous such that a situation of more than one scattering screen is not expected to be rare. In particular, double-lens scintillometry should be a useful tool for sources that have a local environment likely hosting a screen in addition to the screens expected closer to Earth, namely pulsars in Supernova remnants like the Crab and B0540-69, the galactic centre magnetar, and extragalactic sources like pulsars in the Magellanic clouds or fast radio bursts (FRBs).

\section*{Acknowledgements}

This work is based on observations with the 100-m telescope of the MPIfR (Max-Planck-Institut für Radioastronomie) at Effelsberg. We thank Alex Kraus for scheduling additional time and the operators of the telescope for executing this as a filler project.

Tim Sprenger is a member of the International Max Planck Research School for Astronomy and Astrophysics at the Universities of Bonn and Cologne.

We thank Daniel Baker, Marten van Kerkwijk and Rik van Lieshout for helpful discussions during various \emph{theta-theta meetings}.

\section*{Data Availability}

Data will be shared upon reasonable request to the corresponding author. Furthermore, the data is planned to be made publicly available by means yet to be decided.



\bibliographystyle{mnras}
\bibliography{B1508}

\begin{thebibliography}{}
\makeatletter
\relax
\def\mn@urlcharsother{\let\do\@makeother \do\$\do\&\do\#\do\^\do\_\do\%\do\~}
\def\mn@doi{\begingroup\mn@urlcharsother \@ifnextchar [ {\mn@doi@}
  {\mn@doi@[]}}
\def\mn@doi@[#1]#2{\def\@tempa{#1}\ifx\@tempa\@empty \href
  {http://dx.doi.org/#2} {doi:#2}\else \href {http://dx.doi.org/#2} {#1}\fi
  \endgroup}
\def\mn@eprint#1#2{\mn@eprint@#1:#2::\@nil}
\def\mn@eprint@arXiv#1{\href {http://arxiv.org/abs/#1} {{\tt arXiv:#1}}}
\def\mn@eprint@dblp#1{\href {http://dblp.uni-trier.de/rec/bibtex/#1.xml}
  {dblp:#1}}
\def\mn@eprint@#1:#2:#3:#4\@nil{\def\@tempa {#1}\def\@tempb {#2}\def\@tempc
  {#3}\ifx \@tempc \@empty \let \@tempc \@tempb \let \@tempb \@tempa \fi \ifx
  \@tempb \@empty \def\@tempb {arXiv}\fi \@ifundefined
  {mn@eprint@\@tempb}{\@tempb:\@tempc}{\expandafter \expandafter \csname
  mn@eprint@\@tempb\endcsname \expandafter{\@tempc}}}

\bibitem[\protect\citeauthoryear{{Armstrong} \& {Rickett}}{{Armstrong} \&
  {Rickett}}{1981}]{1981MNRAS.194..623A}
{Armstrong} J.~W.,  {Rickett} B.~J.,  1981, \mn@doi [\mnras]
  {10.1093/mnras/194.3.623}, \href
  {https://ui.adsabs.harvard.edu/abs/1981MNRAS.194..623A} {194, 623}

\bibitem[\protect\citeauthoryear{{Astropy Collaboration} et~al.,}{{Astropy
  Collaboration} et~al.}{2013}]{2013A&A...558A..33A}
{Astropy Collaboration} et~al., 2013, \mn@doi [\aap]
  {10.1051/0004-6361/201322068}, \href
  {https://ui.adsabs.harvard.edu/abs/2013A&A...558A..33A} {558, A33}

\bibitem[\protect\citeauthoryear{{Astropy Collaboration} et~al.,}{{Astropy
  Collaboration} et~al.}{2018}]{2018AJ....156..123A}
{Astropy Collaboration} et~al., 2018, \mn@doi [\aj] {10.3847/1538-3881/aabc4f},
  \href {https://ui.adsabs.harvard.edu/abs/2018AJ....156..123A} {156, 123}

\bibitem[\protect\citeauthoryear{{Baker}, {Brisken}, {van Kerkwijk}, {Main},
  {Pen}, {Sprenger}  \& {Wucknitz}}{{Baker} et~al.}{2022}]{2022MNRAS.510.4573B}
{Baker} D.,  {Brisken} W.,  {van Kerkwijk} M.~H.,  {Main} R.,  {Pen} U.-L.,
  {Sprenger} T.,   {Wucknitz} O.,  2022, \mn@doi [\mnras]
  {10.1093/mnras/stab3599}, \href
  {https://ui.adsabs.harvard.edu/abs/2022MNRAS.510.4573B} {510, 4573}

\bibitem[\protect\citeauthoryear{{Bansal}, {Taylor}, {Stovall}  \&
  {Dowell}}{{Bansal} et~al.}{2020}]{2020ApJ...892...26B}
{Bansal} K.,  {Taylor} G.~B.,  {Stovall} K.,   {Dowell} J.,  2020, \mn@doi
  [\apj] {10.3847/1538-4357/ab76bc}, \href
  {https://ui.adsabs.harvard.edu/abs/2020ApJ...892...26B} {892, 26}

\bibitem[\protect\citeauthoryear{{Bhat}, {Rao}  \& {Gupta}}{{Bhat}
  et~al.}{1999a}]{1999ApJS..121..483B}
{Bhat} N.~D.~R.,  {Rao} A.~P.,   {Gupta} Y.,  1999a, \mn@doi [\apjs]
  {10.1086/313198}, \href
  {https://ui.adsabs.harvard.edu/abs/1999ApJS..121..483B} {121, 483}

\bibitem[\protect\citeauthoryear{{Bhat}, {Gupta}  \& {Rao}}{{Bhat}
  et~al.}{1999b}]{1999ApJ...514..249B}
{Bhat} N.~D.~R.,  {Gupta} Y.,   {Rao} A.~P.,  1999b, \mn@doi [\apj]
  {10.1086/306919}, \href
  {https://ui.adsabs.harvard.edu/abs/1999ApJ...514..249B} {514, 249}

\bibitem[\protect\citeauthoryear{{Bhat}, {Rao}  \& {Gupta}}{{Bhat}
  et~al.}{1999c}]{1999ApJ...514..272B}
{Bhat} N.~D.~R.,  {Rao} A.~P.,   {Gupta} Y.,  1999c, \mn@doi [\apj]
  {10.1086/306920}, \href
  {https://ui.adsabs.harvard.edu/abs/1999ApJ...514..272B} {514, 272}

\bibitem[\protect\citeauthoryear{{Brisken}, {Macquart}, {Gao}, {Rickett},
  {Coles}, {Deller}, {Tingay}  \& {West}}{{Brisken}
  et~al.}{2010}]{2010ApJ...708..232B}
{Brisken} W.~F.,  {Macquart} J.~P.,  {Gao} J.~J.,  {Rickett} B.~J.,  {Coles}
  W.~A.,  {Deller} A.~T.,  {Tingay} S.~J.,   {West} C.~J.,  2010, \mn@doi
  [\apj] {10.1088/0004-637X/708/1/232}, \href
  {https://ui.adsabs.harvard.edu/abs/2010ApJ...708..232B} {708, 232}

\bibitem[\protect\citeauthoryear{{Chatterjee} et~al.,}{{Chatterjee}
  et~al.}{2005}]{2005ApJ...630L..61C}
{Chatterjee} S.,  et~al., 2005, \mn@doi [\apjl] {10.1086/491701}, \href
  {https://ui.adsabs.harvard.edu/abs/2005ApJ...630L..61C} {630, L61}

\bibitem[\protect\citeauthoryear{{Chatterjee} et~al.,}{{Chatterjee}
  et~al.}{2009}]{2009ApJ...698..250C}
{Chatterjee} S.,  et~al., 2009, \mn@doi [\apj] {10.1088/0004-637X/698/1/250},
  \href {https://ui.adsabs.harvard.edu/abs/2009ApJ...698..250C} {698, 250}

\bibitem[\protect\citeauthoryear{{Cordes} \& {Rickett}}{{Cordes} \&
  {Rickett}}{1998}]{1998ApJ...507..846C}
{Cordes} J.~M.,  {Rickett} B.~J.,  1998, \mn@doi [\apj] {10.1086/306358}, \href
  {https://ui.adsabs.harvard.edu/abs/1998ApJ...507..846C} {507, 846}

\bibitem[\protect\citeauthoryear{{Cordes}, {Bhat}, {Hankins}, {McLaughlin}  \&
  {Kern}}{{Cordes} et~al.}{2004}]{2004ApJ...612..375C}
{Cordes} J.~M.,  {Bhat} N.~D.~R.,  {Hankins} T.~H.,  {McLaughlin} M.~A.,
  {Kern} J.,  2004, \mn@doi [\apj] {10.1086/422495}, \href
  {https://ui.adsabs.harvard.edu/abs/2004ApJ...612..375C} {612, 375}

\bibitem[\protect\citeauthoryear{{Cordes}, {Rickett}, {Stinebring}  \&
  {Coles}}{{Cordes} et~al.}{2006}]{2006ApJ...637..346C}
{Cordes} J.~M.,  {Rickett} B.~J.,  {Stinebring} D.~R.,   {Coles} W.~A.,  2006,
  \mn@doi [\apj] {10.1086/498332}, \href
  {https://ui.adsabs.harvard.edu/abs/2006ApJ...637..346C} {637, 346}

\bibitem[\protect\citeauthoryear{{Demorest}}{{Demorest}}{2011}]{2011MNRAS.416.2821D}
{Demorest} P.~B.,  2011, \mn@doi [\mnras] {10.1111/j.1365-2966.2011.19230.x},
  \href {https://ui.adsabs.harvard.edu/abs/2011MNRAS.416.2821D} {416, 2821}

\bibitem[\protect\citeauthoryear{{Dolch} et~al.,}{{Dolch}
  et~al.}{2021}]{2021ApJ...913...98D}
{Dolch} T.,  et~al., 2021, \mn@doi [\apj] {10.3847/1538-4357/abf48b}, \href
  {https://ui.adsabs.harvard.edu/abs/2021ApJ...913...98D} {913, 98}

\bibitem[\protect\citeauthoryear{{Er}, {Wagner}  \& {Mao}}{{Er}
  et~al.}{2022}]{2022MNRAS.509.5872E}
{Er} X.,  {Wagner} J.,   {Mao} S.,  2022, \mn@doi [\mnras]
  {10.1093/mnras/stab3278}, \href
  {https://ui.adsabs.harvard.edu/abs/2022MNRAS.509.5872E} {509, 5872}

\bibitem[\protect\citeauthoryear{{Esamdin}, {Zhou}  \& {Wu}}{{Esamdin}
  et~al.}{2004}]{2004A&A...425..949E}
{Esamdin} A.,  {Zhou} A.~Z.,   {Wu} X.~J.,  2004, \mn@doi [\aap]
  {10.1051/0004-6361:20040490}, \href
  {https://ui.adsabs.harvard.edu/abs/2004A&A...425..949E} {425, 949}

\bibitem[\protect\citeauthoryear{{Feldbrugge}}{{Feldbrugge}}{2020}]{2020arXiv201003089F}
{Feldbrugge} J.,  2020, arXiv e-prints, \href
  {https://ui.adsabs.harvard.edu/abs/2020arXiv201003089F} {p. arXiv:2010.03089}

\bibitem[\protect\citeauthoryear{{Foreman-Mackey}, {Hogg}, {Lang}  \&
  {Goodman}}{{Foreman-Mackey} et~al.}{2013}]{2013PASP..125..306F}
{Foreman-Mackey} D.,  {Hogg} D.~W.,  {Lang} D.,   {Goodman} J.,  2013, \mn@doi
  [\pasp] {10.1086/670067}, \href
  {https://ui.adsabs.harvard.edu/abs/2013PASP..125..306F} {125, 306}

\bibitem[\protect\citeauthoryear{{Gaia Collaboration} et~al.,}{{Gaia
  Collaboration} et~al.}{2018}]{2018A&A...616A..11G}
{Gaia Collaboration} et~al., 2018, \mn@doi [\aap]
  {10.1051/0004-6361/201832865}, \href
  {https://ui.adsabs.harvard.edu/abs/2018A&A...616A..11G} {616, A11}

\bibitem[\protect\citeauthoryear{{Geyer} et~al.,}{{Geyer}
  et~al.}{2021}]{2021MNRAS.505.4468G}
{Geyer} M.,  et~al., 2021, \mn@doi [\mnras] {10.1093/mnras/stab1501}, \href
  {https://ui.adsabs.harvard.edu/abs/2021MNRAS.505.4468G} {505, 4468}

\bibitem[\protect\citeauthoryear{{Gvaramadze}, {Gualandris}  \& {Portegies
  Zwart}}{{Gvaramadze} et~al.}{2008}]{2008MNRAS.385..929G}
{Gvaramadze} V.~V.,  {Gualandris} A.,   {Portegies Zwart} S.,  2008, \mn@doi
  [\mnras] {10.1111/j.1365-2966.2008.12884.x}, \href
  {https://ui.adsabs.harvard.edu/abs/2008MNRAS.385..929G} {385, 929}

\bibitem[\protect\citeauthoryear{{Gwinn}}{{Gwinn}}{2019}]{2019MNRAS.486.2809G}
{Gwinn} C.~R.,  2019, \mn@doi [\mnras] {10.1093/mnras/stz894}, \href
  {https://ui.adsabs.harvard.edu/abs/2019MNRAS.486.2809G} {486, 2809}

\bibitem[\protect\citeauthoryear{{Gwinn} \& {Sosenko}}{{Gwinn} \&
  {Sosenko}}{2019}]{2019MNRAS.489.3692G}
{Gwinn} C.~R.,  {Sosenko} E.~B.,  2019, \mn@doi [\mnras]
  {10.1093/mnras/stz2364}, \href
  {https://ui.adsabs.harvard.edu/abs/2019MNRAS.489.3692G} {489, 3692}

\bibitem[\protect\citeauthoryear{{Gwinn}, {Britton}, {Reynolds}, {Jauncey},
  {King}, {McCulloch}, {Lovell}  \& {Preston}}{{Gwinn}
  et~al.}{1998}]{1998ApJ...505..928G}
{Gwinn} C.~R.,  {Britton} M.~C.,  {Reynolds} J.~E.,  {Jauncey} D.~L.,  {King}
  E.~A.,  {McCulloch} P.~M.,  {Lovell} J.~E.~J.,   {Preston} R.~A.,  1998,
  \mn@doi [\apj] {10.1086/306178}, \href
  {https://ui.adsabs.harvard.edu/abs/1998ApJ...505..928G} {505, 928}

\bibitem[\protect\citeauthoryear{{Hill}, {Stinebring}, {Asplund}, {Berwick},
  {Everett}  \& {Hinkel}}{{Hill} et~al.}{2005}]{2005ApJ...619L.171H}
{Hill} A.~S.,  {Stinebring} D.~R.,  {Asplund} C.~T.,  {Berwick} D.~E.,
  {Everett} W.~B.,   {Hinkel} N.~R.,  2005, \mn@doi [\apjl] {10.1086/428347},
  \href {https://ui.adsabs.harvard.edu/abs/2005ApJ...619L.171H} {619, L171}

\bibitem[\protect\citeauthoryear{{Hobbs}, {Lyne}, {Kramer}, {Martin}  \&
  {Jordan}}{{Hobbs} et~al.}{2004}]{2004MNRAS.353.1311H}
{Hobbs} G.,  {Lyne} A.~G.,  {Kramer} M.,  {Martin} C.~E.,   {Jordan} C.,  2004,
  \mn@doi [\mnras] {10.1111/j.1365-2966.2004.08157.x}, \href
  {https://ui.adsabs.harvard.edu/abs/2004MNRAS.353.1311H} {353, 1311}

\bibitem[\protect\citeauthoryear{{Huguenin}, {Taylor}, {Goad}, {Hartai},
  {Orsten}  \& {Rodman}}{{Huguenin} et~al.}{1968}]{1968Natur.219..576H}
{Huguenin} G.~R.,  {Taylor} J.~H.,  {Goad} L.~E.,  {Hartai} A.,  {Orsten}
  G.~S.~F.,   {Rodman} A.~K.,  1968, \mn@doi [\nat] {10.1038/219576a0}, \href
  {https://ui.adsabs.harvard.edu/abs/1968Natur.219..576H} {219, 576}

\bibitem[\protect\citeauthoryear{{Kaspi} \& {Stinebring}}{{Kaspi} \&
  {Stinebring}}{1992}]{1992ApJ...392..530K}
{Kaspi} V.~M.,  {Stinebring} D.~R.,  1992, \mn@doi [\apj] {10.1086/171454},
  \href {https://ui.adsabs.harvard.edu/abs/1992ApJ...392..530K} {392, 530}

\bibitem[\protect\citeauthoryear{{Kondratiev}, {Popov}, {Soglasnov}  \&
  {Kostyuk}}{{Kondratiev} et~al.}{2001}]{2001Ap&SS.278...43K}
{Kondratiev} V.~I.,  {Popov} M.~V.,  {Soglasnov} V.~A.,   {Kostyuk} S.~V.,
  2001, \mn@doi [\apss] {10.1023/A:1013177621936}, \href
  {https://ui.adsabs.harvard.edu/abs/2001Ap&SS.278...43K} {278, 43}

\bibitem[\protect\citeauthoryear{{Liu}, {Pen}, {Macquart}, {Brisken}  \&
  {Deller}}{{Liu} et~al.}{2016}]{2016MNRAS.458.1289L}
{Liu} S.,  {Pen} U.-L.,  {Macquart} J.~P.,  {Brisken} W.,   {Deller} A.,  2016,
  \mn@doi [\mnras] {10.1093/mnras/stw314}, \href
  {https://ui.adsabs.harvard.edu/abs/2016MNRAS.458.1289L} {458, 1289}

\bibitem[\protect\citeauthoryear{{Liu} et~al.,}{{Liu}
  et~al.}{2017}]{2017ApJ...834...33L}
{Liu} W.,  et~al., 2017, \mn@doi [\apj] {10.3847/1538-4357/834/1/33}, \href
  {https://ui.adsabs.harvard.edu/abs/2017ApJ...834...33L} {834, 33}

\bibitem[\protect\citeauthoryear{{Lyne} \& {Smith}}{{Lyne} \&
  {Smith}}{1982}]{1982Natur.298..825L}
{Lyne} A.~G.,  {Smith} F.~G.,  1982, \mn@doi [\nat] {10.1038/298825a0}, \href
  {https://ui.adsabs.harvard.edu/abs/1982Natur.298..825L} {298, 825}

\bibitem[\protect\citeauthoryear{{Main}, {van Kerkwijk}, {Pen}, {Mahajan}  \&
  {Vanderlinde}}{{Main} et~al.}{2017}]{2017ApJ...840L..15M}
{Main} R.,  {van Kerkwijk} M.,  {Pen} U.-L.,  {Mahajan} N.,   {Vanderlinde} K.,
   2017, \mn@doi [\apjl] {10.3847/2041-8213/aa6f03}, \href
  {https://ui.adsabs.harvard.edu/abs/2017ApJ...840L..15M} {840, L15}

\bibitem[\protect\citeauthoryear{{Main} et~al.,}{{Main}
  et~al.}{2020}]{2020MNRAS.499.1468M}
{Main} R.~A.,  et~al., 2020, \mn@doi [\mnras] {10.1093/mnras/staa2955}, \href
  {https://ui.adsabs.harvard.edu/abs/2020MNRAS.499.1468M} {499, 1468}

\bibitem[\protect\citeauthoryear{{Mall} et~al.,}{{Mall}
  et~al.}{2022}]{2022MNRAS.511.1104M}
{Mall} G.,  et~al., 2022, \mn@doi [\mnras] {10.1093/mnras/stac096}, \href
  {https://ui.adsabs.harvard.edu/abs/2022MNRAS.511.1104M} {511, 1104}

\bibitem[\protect\citeauthoryear{{Marthi} et~al.,}{{Marthi}
  et~al.}{2021}]{2021MNRAS.506.5160M}
{Marthi} V.~R.,  et~al., 2021, \mn@doi [\mnras] {10.1093/mnras/stab1970}, \href
  {https://ui.adsabs.harvard.edu/abs/2021MNRAS.506.5160M} {506, 5160}

\bibitem[\protect\citeauthoryear{{McKee}, {Lyne}, {Stappers}, {Bassa}  \&
  {Jordan}}{{McKee} et~al.}{2018}]{2018MNRAS.479.4216M}
{McKee} J.~W.,  {Lyne} A.~G.,  {Stappers} B.~W.,  {Bassa} C.~G.,   {Jordan}
  C.~A.,  2018, \mn@doi [\mnras] {10.1093/mnras/sty1727}, \href
  {https://ui.adsabs.harvard.edu/abs/2018MNRAS.479.4216M} {479, 4216}

\bibitem[\protect\citeauthoryear{{McKee}, {Zhu}, {Stinebring}  \&
  {Cordes}}{{McKee} et~al.}{2022}]{2022ApJ...927...99M}
{McKee} J.~W.,  {Zhu} H.,  {Stinebring} D.~R.,   {Cordes} J.~M.,  2022, \mn@doi
  [\apj] {10.3847/1538-4357/ac460b}, \href
  {https://ui.adsabs.harvard.edu/abs/2022ApJ...927...99M} {927, 99}

\bibitem[\protect\citeauthoryear{{Pen} \& {Levin}}{{Pen} \&
  {Levin}}{2014}]{2014MNRAS.442.3338P}
{Pen} U.-L.,  {Levin} Y.,  2014, \mn@doi [\mnras] {10.1093/mnras/stu1020},
  \href {https://ui.adsabs.harvard.edu/abs/2014MNRAS.442.3338P} {442, 3338}

\bibitem[\protect\citeauthoryear{{Putney} \& {Stinebring}}{{Putney} \&
  {Stinebring}}{2006}]{2006ChJAS...6b.233P}
{Putney} M.~L.,  {Stinebring} D.~R.,  2006, Chinese Journal of Astronomy and
  Astrophysics Supplement, \href
  {https://ui.adsabs.harvard.edu/abs/2006ChJAS...6b.233P} {6, 233}

\bibitem[\protect\citeauthoryear{{Reardon} et~al.,}{{Reardon}
  et~al.}{2020}]{2020ApJ...904..104R}
{Reardon} D.~J.,  et~al., 2020, \mn@doi [\apj] {10.3847/1538-4357/abbd40},
  \href {https://ui.adsabs.harvard.edu/abs/2020ApJ...904..104R} {904, 104}

\bibitem[\protect\citeauthoryear{{Rickett}}{{Rickett}}{1970}]{1970MNRAS.150...67R}
{Rickett} B.~J.,  1970, \mn@doi [\mnras] {10.1093/mnras/150.1.67}, \href
  {https://ui.adsabs.harvard.edu/abs/1970MNRAS.150...67R} {150, 67}

\bibitem[\protect\citeauthoryear{{Rickett}}{{Rickett}}{1990}]{1990ARA&A..28..561R}
{Rickett} B.~J.,  1990, \mn@doi [\araa] {10.1146/annurev.aa.28.090190.003021},
  \href {https://ui.adsabs.harvard.edu/abs/1990ARA&A..28..561R} {28, 561}

\bibitem[\protect\citeauthoryear{{Rickett}, {Stinebring}, {Zhu}  \&
  {Minter}}{{Rickett} et~al.}{2021}]{2021ApJ...907...49R}
{Rickett} B.~J.,  {Stinebring} D.~R.,  {Zhu} H.,   {Minter} A.~H.,  2021,
  \mn@doi [\apj] {10.3847/1538-4357/abc9bc}, \href
  {https://ui.adsabs.harvard.edu/abs/2021ApJ...907...49R} {907, 49}

\bibitem[\protect\citeauthoryear{{Shi}}{{Shi}}{2021}]{2021MNRAS.508..125S}
{Shi} X.,  2021, \mn@doi [\mnras] {10.1093/mnras/stab2522}, \href
  {https://ui.adsabs.harvard.edu/abs/2021MNRAS.508..125S} {508, 125}

\bibitem[\protect\citeauthoryear{{Simard} \& {Pen}}{{Simard} \&
  {Pen}}{2018}]{2018MNRAS.478..983S}
{Simard} D.,  {Pen} U.-L.,  2018, \mn@doi [\mnras] {10.1093/mnras/sty1140},
  \href {https://ui.adsabs.harvard.edu/abs/2018MNRAS.478..983S} {478, 983}

\bibitem[\protect\citeauthoryear{{Smirnova} et~al.,}{{Smirnova}
  et~al.}{2020}]{2020MNRAS.496.5149S}
{Smirnova} T.~V.,  et~al., 2020, \mn@doi [\mnras] {10.1093/mnras/staa1839},
  \href {https://ui.adsabs.harvard.edu/abs/2020MNRAS.496.5149S} {496, 5149}

\bibitem[\protect\citeauthoryear{{Smith} \& {Wright}}{{Smith} \&
  {Wright}}{1985}]{1985MNRAS.214...97S}
{Smith} F.~G.,  {Wright} N.~C.,  1985, \mn@doi [\mnras]
  {10.1093/mnras/214.2.97}, \href
  {https://ui.adsabs.harvard.edu/abs/1985MNRAS.214...97S} {214, 97}

\bibitem[\protect\citeauthoryear{{Sprenger}, {Wucknitz}, {Main}, {Baker}  \&
  {Brisken}}{{Sprenger} et~al.}{2021}]{2021MNRAS.500.1114S}
{Sprenger} T.,  {Wucknitz} O.,  {Main} R.,  {Baker} D.,   {Brisken} W.,  2021,
  \mn@doi [\mnras] {10.1093/mnras/staa3353}, \href
  {https://ui.adsabs.harvard.edu/abs/2021MNRAS.500.1114S} {500, 1114}

\bibitem[\protect\citeauthoryear{{Stinebring}}{{Stinebring}}{2007a}]{2007A&AT...26..517S}
{Stinebring} D.,  2007a, \mn@doi [Astronomical and Astrophysical Transactions]
  {10.1080/10556790701614862}, \href
  {https://ui.adsabs.harvard.edu/abs/2007A&AT...26..517S} {26, 517}

\bibitem[\protect\citeauthoryear{{Stinebring}}{{Stinebring}}{2007b}]{2007ASPC..365..254S}
{Stinebring} D.,  2007b, in {Haverkorn} M.,  {Goss} W.~M.,  eds,  Astronomical
  Society of the Pacific Conference Series Vol. 365, SINS - Small Ionized and
  Neutral Structures in the Diffuse Interstellar Medium. p.~254

\bibitem[\protect\citeauthoryear{{Stinebring}, {McLaughlin}, {Cordes},
  {Becker}, {Goodman}, {Kramer}, {Sheckard}  \& {Smith}}{{Stinebring}
  et~al.}{2001}]{2001ApJ...549L..97S}
{Stinebring} D.~R.,  {McLaughlin} M.~A.,  {Cordes} J.~M.,  {Becker} K.~M.,
  {Goodman} J.~E.~E.,  {Kramer} M.~A.,  {Sheckard} J.~L.,   {Smith} C.~T.,
  2001, \mn@doi [\apjl] {10.1086/319133}, \href
  {https://ui.adsabs.harvard.edu/abs/2001ApJ...549L..97S} {549, L97}

\bibitem[\protect\citeauthoryear{{Stovall} et~al.,}{{Stovall}
  et~al.}{2015}]{2015ApJ...808..156S}
{Stovall} K.,  et~al., 2015, \mn@doi [\apj] {10.1088/0004-637X/808/2/156},
  \href {https://ui.adsabs.harvard.edu/abs/2015ApJ...808..156S} {808, 156}

\bibitem[\protect\citeauthoryear{{Walker}, {Melrose}, {Stinebring}  \&
  {Zhang}}{{Walker} et~al.}{2004}]{2004MNRAS.354...43W}
{Walker} M.~A.,  {Melrose} D.~B.,  {Stinebring} D.~R.,   {Zhang} C.~M.,  2004,
  \mn@doi [\mnras] {10.1111/j.1365-2966.2004.08159.x}, \href
  {https://ui.adsabs.harvard.edu/abs/2004MNRAS.354...43W} {354, 43}

\bibitem[\protect\citeauthoryear{{Walker}, {Koopmans}, {Stinebring}  \& {van
  Straten}}{{Walker} et~al.}{2008}]{2008MNRAS.388.1214W}
{Walker} M.~A.,  {Koopmans} L.~V.~E.,  {Stinebring} D.~R.,   {van Straten} W.,
  2008, \mn@doi [\mnras] {10.1111/j.1365-2966.2008.13452.x}, \href
  {https://ui.adsabs.harvard.edu/abs/2008MNRAS.388.1214W} {388, 1214}

\bibitem[\protect\citeauthoryear{{Walker}, {Demorest}  \& {van
  Straten}}{{Walker} et~al.}{2013}]{2013ApJ...779...99W}
{Walker} M.~A.,  {Demorest} P.~B.,   {van Straten} W.,  2013, \mn@doi [\apj]
  {10.1088/0004-637X/779/2/99}, \href
  {https://ui.adsabs.harvard.edu/abs/2013ApJ...779...99W} {779, 99}

\bibitem[\protect\citeauthoryear{{Walker}, {Tuntsov}, {Bignall}, {Reynolds},
  {Bannister}, {Johnston}, {Stevens}  \& {Ravi}}{{Walker}
  et~al.}{2017}]{2017ApJ...843...15W}
{Walker} M.~A.,  {Tuntsov} A.~V.,  {Bignall} H.,  {Reynolds} C.,  {Bannister}
  K.~W.,  {Johnston} S.,  {Stevens} J.,   {Ravi} V.,  2017, \mn@doi [\apj]
  {10.3847/1538-4357/aa705c}, \href
  {https://ui.adsabs.harvard.edu/abs/2017ApJ...843...15W} {843, 15}

\bibitem[\protect\citeauthoryear{{Wang} et~al.,}{{Wang}
  et~al.}{2018a}]{2018yCat..36180186W}
{Wang} P.~F.,  et~al., 2018a, VizieR Online Data Catalog, \href
  {https://ui.adsabs.harvard.edu/abs/2018yCat..36180186W} {pp J/A+A/618/A186}

\bibitem[\protect\citeauthoryear{{Wang} et~al.,}{{Wang}
  et~al.}{2018b}]{2018A&A...618A.186W}
{Wang} P.~F.,  et~al., 2018b, \mn@doi [\aap] {10.1051/0004-6361/201833215},
  \href {https://ui.adsabs.harvard.edu/abs/2018A&A...618A.186W} {618, A186}

\bibitem[\protect\citeauthoryear{{Wucknitz}}{{Wucknitz}}{2018}]{2018evn..confE..17W}
{Wucknitz} O.,  2018, in 14th European VLBI Network Symposium \& Users Meeting
  (EVN 2018). p.~17

\bibitem[\protect\citeauthoryear{{Yao} et~al.,}{{Yao}
  et~al.}{2021}]{2021NatAs...5..788Y}
{Yao} J.,  et~al., 2021, \mn@doi [Nature Astronomy]
  {10.1038/s41550-021-01360-w}, \href
  {https://ui.adsabs.harvard.edu/abs/2021NatAs...5..788Y} {5, 788}

\bibitem[\protect\citeauthoryear{{van Straten} \& {Bailes}}{{van Straten} \&
  {Bailes}}{2011}]{2011PASA...28....1V}
{van Straten} W.,  {Bailes} M.,  2011, \mn@doi [\pasa] {10.1071/AS10021}, \href
  {https://ui.adsabs.harvard.edu/abs/2011PASA...28....1V} {28, 1}

\bibitem[\protect\citeauthoryear{{van Straten}, {Demorest}  \& {Oslowski}}{{van
  Straten} et~al.}{2012}]{2012AR&T....9..237V}
{van Straten} W.,  {Demorest} P.,   {Oslowski} S.,  2012, Astronomical Research
  and Technology, \href {https://ui.adsabs.harvard.edu/abs/2012AR&T....9..237V}
  {9, 237}

\makeatother
\end{thebibliography}



\appendix

\section{Derivation of a two-screen theory}
\label{Sec:Derivations}

We derive the theory of a two-screen model starting from Kirchhoff diffraction, which gives the complex electric field at the destination plane by an integration over all paths from the source plane:
\begin{align}
    E_\text{destination}(\bm{p}) = \frac{i}{\lambda} \int \der^2 s \frac{\text{e}^{2\pi i\frac{d(\bm{p},\bm{s})}{\lambda}}}{d(\bm{p},\bm{s})} E_\text{source}(\bm{s}) \, .
\end{align}
Here, $\bm{p}$ and $\bm{s}$ are the two-dimensional coordinates on the destination and source plane and $d(\bm{p},\bm{s})$ is the distance between these locations on the two screens.

For scintillation studies, we are not interested in the prefactor. Additionally, we regard screens that are much farther apart than their extent, which means we can safely approximate the denominator as constant and expand the phase term to leading order around $\bm{p}=\bm{s}=0$. After dropping all prefactors, we arrive at
\begin{align}
    E_\text{destination}(\bm{p}) = \int \der^2 s \,\text{e}^{ i\frac{\pi\nu}{c} \frac{\left( \bm{p}-\bm{s} \right)^2}{D_{p,s}}} E_\text{source}(\bm{s}) \, .
\end{align}
Above, $D_{p,s}$ is the perpendicular distance between the screens. This formula is modular. Hence we can replace the source term by another Kirchhoff integral describing a diffraction screen as an additional plane. Because of the huge distances, all screens can be treated as thin (two-dimensional) and perpendicular to the line of sight.

The electric field in the source plane identified with the pulsar will be treated as $E_\text{source}=1$ now because we do not know the intrinsic electric field other than that it is sufficiently point-like. If the screen structure would be known however, this would be the starting point to infer information through interstellar interferometry.

A diffraction screen induces an extra phase shift on the passing radiation. If caused by the presence of free electrons, which is the canonical explanation for pulsar scintillation, this phase shift can be related to the dispersion measure (DM) of this path of propagation crossing the screen at $\bm{x}$:
\begin{align}
    \phi(\bm{x}) = -\frac{1}{4\pi\epsilon_0}\,\frac{e^2}{m_e c}\,\frac{\text{DM}(\bm{x})}{\nu} \, ,
\end{align}
where $e$ is the elementary charge and $m_e$ is the mass of the electron.

\begin{figure}
 \includegraphics[width=\columnwidth]{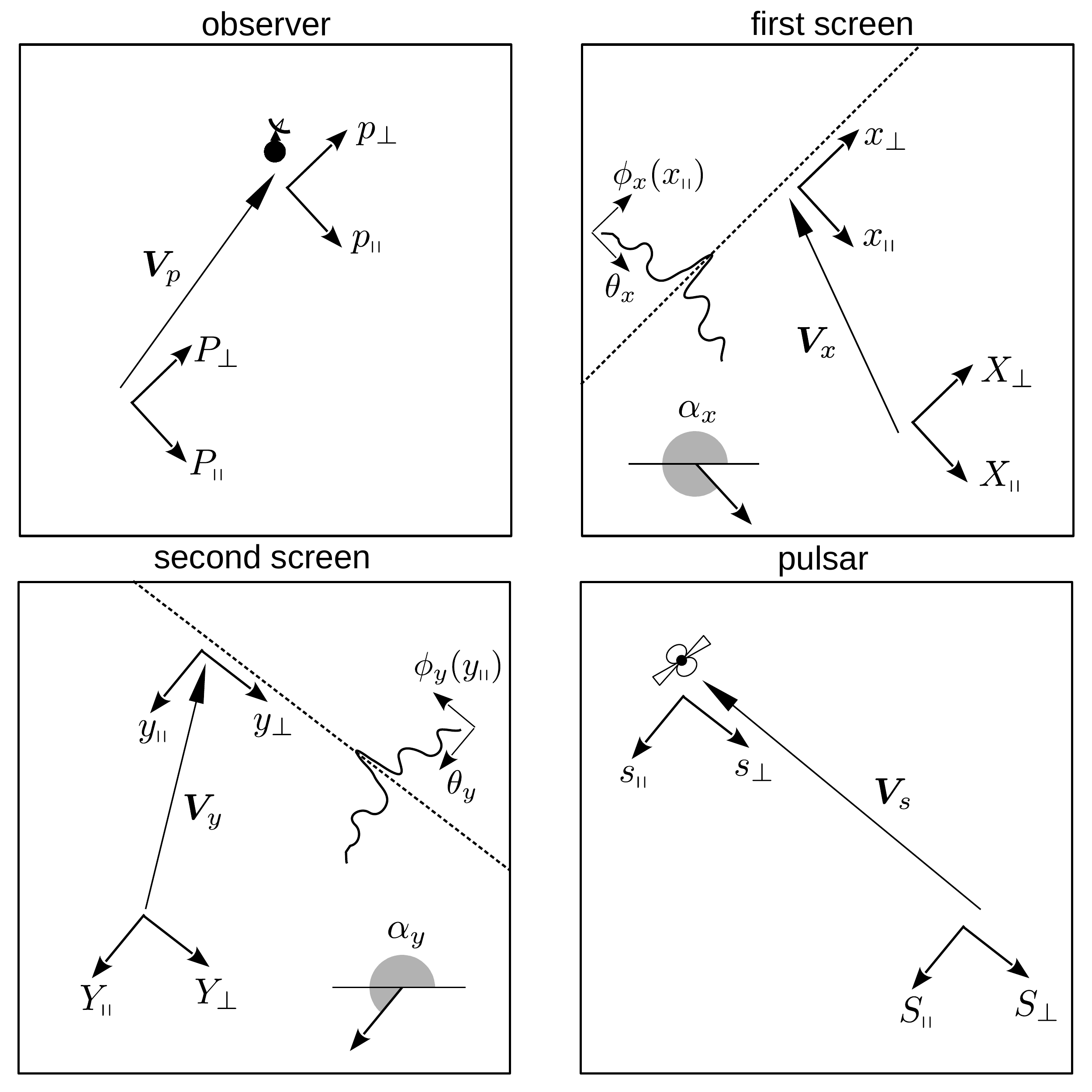}
 \vspace{-3mm}
 \caption{
The notation used in each of the four planes of a double-screen situation. Upper case symbols are coordinates fixed within the planes with respect to the initial line of sight, while lower case symbols are coordinates that are fixed to the moving object. The coordinate systems are aligned to the axes of anisotropy on the screens. In the case of Earth and pulsar, they are aligned to the closer respective screen, which eases derivations. Compare \cref{Fig:DoubleScreenVisualization} for the notation of distances between the planes.}
 \label{Fig:DoubleScreenNotation}
\end{figure}

Even with the above approximations being valid, the screen is still likely a very large structure. Hence it can be approximated as being frozen in time, which makes it convenient to separate the frame velocity of the screens from coordinates within those screens, i.e.~$\bm{X} = \bm{x} + V_x t$. For the remaining part of this section we will use upper case letters for time dependent variables and lower case letters for coordinates on the frozen comoving planes. A sketch illustrating these notations is given in \cref{Fig:DoubleScreenNotation}.

In its most general form, the phase inflicted on a path of propagation with respect to the unperturbed line of sight is given by
\begin{align}
 \Phi = \sum_{n=1}^{N}\phi_n(\bm{x}_n) + \sum_{n=1}^{N+1} \frac{\pi\nu}{c} \frac{\left( \bm{x}_n - \bm{x}_{n-1} +(\bm{V}_n-\bm{V}_{n-1})t \right)^2}{D_{n-1,n}} \label{Eq:2ScreenGeneral}
\end{align}
and the electric field at the observer can be computed via the integral
\begin{align}
    E(\bm{x}_0) = \int \left(\prod_{n=1}^{N+1}\der^2 x_{n}\right) \text{e}^{i\Phi} \, .
\end{align}
Here, $N$ is the total number of diffraction screens and $\bm{x}_n$ ,$V_n$, and $\phi_n$ are the coordinate on the nth screen, its velocity, and its inflicted phase respectively. $D_{n-1,n}$ is the distance between the n'th screen and the screen before it from the view of the observer. Screen 0 is the observers plane denoted with $p$ in the following while screen N+1 is the source plane denoted with $s$.

The model we are interested in has only two screens. The one closer to the observer will be denoted with $x$ and the other one with $y$. Furthermore, we assume these screens to be one-dimensional, such that they have the same high anisotropy that was assumed in the single screen model because the data shows no strong deviations of features from their main scintillation arc. This means the inflicted dispersive phase only depends on the coordinate parallel to the axis of anisotropy of the screen. Another argument for this model is that the apex of the parabola is always centered at the origin which means --as has been demonstrated by \citet{2021MNRAS.508..125S}-- that the phase structure extends quasi infinitely perpendicular to the anisotropy's axis. Thus, this model is determined by the following phase:
\begin{align}
 \Phi = \phi_x(x_\shortparallel) +\phi_y(y_\shortparallel) + \frac{\pi\nu}{c} \left[
  \frac{\left( \bm{X} - \bm{P} \right)^2}{D_{x}} + \frac{\left( \bm{Y} - \bm{X} \right)^2}{D_{x,y}} + \frac{\left( \bm{S} - \bm{Y} \right)^2}{D_{y,s}}
  \right] \, . 
\end{align}
This expression can be reduced to the relevant coordinates $x_\shortparallel$ and $y_\shortparallel$ by performing the integrals over the perpendicular coordinates $x_\perp$ and $y_\perp$. This amounts to integrals over a quadratic phase. Hence the stationary phase approximation yields the correct remaining phase up to constants and a prefactor, both of which we will again drop. Thus, we insert the following conditions:
\begin{align}
    \frac{\partial\Phi}{\partial x_\perp}  = \frac{\partial \Phi}{\partial y_\perp} = 0 \, .
\end{align}
The result is a quite lengthy expression, so we will employ some abbreviations here. First, we concentrate on the part $\tilde{\Phi}$ that depends on the screen coordinates, such that
\begin{align}
 \Phi = \phi_x(x_\shortparallel) +\phi_y(y_\shortparallel) + \frac{\pi\nu}{c} \left[
  \frac{\bm{P}^2}{D_{x}} + \tilde{\Phi}(\bm{x},\bm{y}) + \frac{\bm{S}^2}{D_{y,s}}
  \right] \, . \label{Eq:Phase_abbr}
\end{align}
Next, we define the terms involving the angles $\alpha_x$ and $\alpha_y$ of the anisotropy axes --the screen axes-- with respect to some common reference frame as
\begin{align}
    \gamma &\equiv \sin(\alpha_x-\alpha_y) \, , \\
    \delta &\equiv \cos(\alpha_x-\alpha_y) \, .
\end{align}
Last, we define coordinates on the source and observer planes as projections onto their respective closest screen according to
\begin{align}
\bm{x}\cdot\bm{p} &\equiv x_\shortparallel p_\shortparallel + x_\perp p_\perp \, , \\
\bm{y}\cdot\bm{s} &\equiv y_\shortparallel s_\shortparallel + y_\perp s_\perp \, .
\end{align}
We define perpendicular here as a vector rotated by $90^\circ$ counterclockwise on the sky, which we define as the positive direction of angles. The perpendicular coordinates are now determined as functions of the parallel coordinates:
\begin{align}
&X_\perp = \nonumber\\ &\frac{ D_x D_{y,s}\gamma\delta X_\shortparallel + D_x D_{x,y}\delta S_\perp - D_x D_{x,s}\gamma Y_\shortparallel + D_{x,s}D_{x,y}P_\perp }{ D_y D_{x,s} - D_x D_{y,s}\delta^2 } \, ,  \label{Eq:x_perp}\\
&Y_\perp = \nonumber\\ &\frac{ -D_x D_{y,s}\gamma\delta Y_\shortparallel + D_{y,s} D_{x,y}\delta P_\perp + D_y D_{y,s}\gamma X_\shortparallel + D_{y}D_{x,y}S_\perp }{ D_y D_{x,s} - D_x D_{y,s}\delta^2 } \, .
\end{align}
Using the above notation, the desired result after integrating the perpendicular coordinates can be written as
\begin{align}
\tilde{\Phi} =&  -2 \frac{1}{D_x}X_\shortparallel P_\shortparallel -2 \frac{1}{D_{y,s}}Y_\shortparallel S_\shortparallel + \frac{1}{ D_y D_{x,s} - D_x D_{y,s}\delta^2 } \times \nonumber\\
&\nonumber \bigg( \frac{D_s D_y}{D_x}X_\shortparallel^2 -2\delta D_s X_\shortparallel Y_\shortparallel + \frac{D_s D_{x,s}}{D_{y,s}}Y_\shortparallel^2 - \frac{D_{y}D_{x,y}}{D_{y,s}}S_\perp^2 \\ &\nonumber -2\delta D_{x,y} P_\perp S_\perp
-\frac{D_{x,y}D_{x,s}}{D_{x}}P_\perp^2 -2\gamma \delta D_{y,s}X_\shortparallel P_\perp  \\ &  -2\gamma D_y X_\shortparallel S_\perp +2\gamma D_{x,s}Y_\shortparallel P_\perp +2\gamma \delta D_x Y_\shortparallel S_\perp \bigg) \, .
\label{Eq:2scr_phitilde}
\end{align}
This phase can now be used to construct effective distances and velocities such that \cref{Eq:Def_fD,Eq:Def_tau} are formally correct in the two screen model. Since we treat the pulsar as a point source and regard only single-station observations, the initial coordinates of pulsar and observer can be chosen to be zero:
\begin{align}
    \bm{P} &= \bm{V}_p t \, , \\
    \bm{S} &= \bm{V}_s t \, .
\end{align}
Because of their definition through a Fourier transform, the delay and Doppler rate can be defined in a general way as derivatives along frequency and time that are evaluated at the initial coordinates:
\begin{align}
    \tau &= \left.\frac{\partial\Phi}{2\pi\partial\nu}\right|_{t=0} \, ,\\
    f_\text{D} &= \left.\frac{\partial\Phi}{2\pi\partial t}\right|_{t=0} \, .
\end{align}
In order to achieve formal similarity to the one-screen result in terms of effective quantities obeying similar relations for delay and Doppler rate and to ease predictions, the geometrical phase can be rewritten as
\begin{align}
 \Phi = \frac{\pi \nu}{c}\Big( D_\text{eff,x}\theta_x^2-2V_{\text{eff},x,\shortparallel}t\theta_x + D_\text{eff,y}\theta_y^2-2V_{\text{eff},y,\shortparallel}t\theta_y  \nonumber\\
 + K t^2 -2 D_\text{mix} \theta_x \theta_y \Big) \, ,
\end{align}
where $K$ is a constant that is irrelevant to further analysis because only the terms varying in angles on the screen can lead to visible interferences. 

For the scintillation arc caused by the angles $\theta_x= x_\shortparallel/D_x$ representing the closer screen, only terms including this angle are relevant because the other terms just represent factors that all angles have in common and thus do not affect the interference leading to the scintillation arc. Also, the initial coordinate on the other screen can be set to zero to get the unperturbed scintillation arc, although its distribution around zero will blur the arc in practise. Therefore, the effective distance and velocity of the closer screen are
\begin{align}
    D_{\text{eff},x} =& \frac{D_s D_y D_x}{ D_y D_{x,s} - D_x D_{y,s}\delta^2 } \, , \\
    V_{\text{eff},x,\shortparallel} =& \big[ -D_s D_y V_{x,\shortparallel} +\delta D_x D_s V_{y,\shortparallel} \nonumber\\
    & +\gamma \delta D_x D_{y,s}V_{p,\perp} +\gamma D_x D_y V_{s,\perp}\big] \nonumber\\
    &/ ~ \big(D_y D_{x,s} - D_x D_{y,s}\delta^2\big) + V_{p,\shortparallel} \, .
\end{align}

Analogously, the effective quantities can be derived for the farther screen with angles $\theta_y= y_\shortparallel/D_y$:
\begin{align}
    D_{\text{eff},y} =& \frac{D_s D_{x,s} D_y^2 / D_{y,s}}{ D_y D_{x,s} - D_x D_{y,s}\delta^2 } \, , \\
    V_{\text{eff},y,\shortparallel} =& \big[ -D_s D_{x,s} D_y/D_{y,s}\, V_{y,\shortparallel} +\delta D_y D_s V_{x,\shortparallel} \nonumber\\
    &  -\gamma D_y D_{x,s}V_{p,\perp} -\gamma\delta D_x D_y V_{s,\perp} \big] \nonumber\\
    & / ~ \big(D_y D_{x,s} - D_x D_{y,s}\delta^2\big) + D_y/D_{y,s}\,V_{s,\shortparallel} \, .
\end{align}
$D_\text{mix}$ is given by:
\begin{align}
    D_\text{mix} = \frac{\delta D_s D_x D_y}{D_y D_{x,s} - D_x D_{y,s}\delta^2} \, .
\end{align}
The mixed term represents an interaction between the two screens that will make a difference in the formal result of the movement of features. The feature movement can be computed as the central line of sight. It can be obtained by simultaneously enforcing the stationary phase condition on both screens and in the absence of dispersive terms:
\begin{align}
 \frac{\partial \Phi}{\partial \theta_x} = 0  ~~~ \text{and} ~~~ \frac{\partial \Phi}{\partial \theta_y} = 0 \, .
\end{align}
As a result, \cref{eq:th0_dot} has to be slightly modified:
\begin{align}
    \dot{\theta}_{x,0} &= \frac{D_{\text{eff},y} V_{\text{eff},x,\shortparallel} + D_\text{mix}V_{\text{eff},y,\shortparallel}}{D_{\text{eff},y} D_{\text{eff},x}-D_\text{mix}^2} \, ,\\
    \dot{\theta}_{y,0} &= \frac{D_{\text{eff},x} V_{\text{eff},y,\shortparallel} + D_\text{mix}V_{\text{eff},x,\shortparallel}}{D_{\text{eff},x} D_{\text{eff},y}-D_\text{mix}^2} \, .
\end{align}

The phenomenon of eigenvector modulation as described in \cref{Sec:Eigenvectors,Fig:Eigenvector_series} can be interpreted as a projection effect. Different lines of sight through the second screen impose different amplifications to the signal. Over time, these sight lines sweep with a single velocity along the structures on the first screen, which results in the observed modulation speed. The moving position $\bm{X}$ of constant amplification on the first screen is given by a straight line from the pulsar through a fixed position $\bm{Y}$ on the second screen and thus satisfies the intercept theorem
\begin{align}
    \frac{\bm{Y}-\bm{X}}{D_{x,y}} = \frac{\bm{S}-\bm{X}}{D_{x,s}} \, .
\end{align}
Because of the total anisotropy, the optical situation is the same for all perpendicular components $Y_\perp$. Thus, this equation defines a line on the first screen whose intersection with the line of images that is actual visible on Earth needs to be computed. The perpendicular coordinate of the visible line is given by \cref{Eq:x_perp}. After aligning the coordinate system with the orientation of the second screen, the equation of the desired location becomes
\begin{align}
    \begin{pmatrix} \delta X_\shortparallel-\gamma X_\perp \\ \gamma X_\shortparallel+\delta X_\perp \end{pmatrix} = \frac{D_{x,s}}{D_{y,s}}\begin{pmatrix} Y_\shortparallel \\ Y_\perp \end{pmatrix} - \frac{D_{x,y}}{D_{y,s}}\begin{pmatrix} S_\shortparallel \\ S_\perp \end{pmatrix} \, .
\end{align}
Since the first component contains no instance of $Y_\perp$, it is sufficient to solve the first component for $X_\shortparallel$. We are interested in the velocity $V_m$ relative to the structure on the phase screen, i.e. we define the solution as
\begin{align}
    X_{\shortparallel,\text{modulation}} = x_\shortparallel + V_{x,\shortparallel} t + V_m t \, .
\end{align}
Taking the derivative with respect to time, we arrive at
\begin{align}
    V_m =& \Big[ D_{s}D_{x,s} V_{y,\shortparallel} + D_x D_{y,s}\gamma\delta V_{s,\perp} + D_{x,s} D_{y,s} \gamma V_{p,\perp} \nonumber\\
    &  + \left( D_x D_{y,s}\delta^2 - D_{x,s}D_y \right)V_{s,\shortparallel} \Big]  / \left( D_{y,s}D_{s}\delta \right) ~ - V_{x,\shortparallel} \, .
\end{align}
Finally, we can translate this result to model-independent units by computing the resulting movement again in a drift rate in $\sqrt{\tau}$:
\begin{align}
    \left(\partial_t \sqrt{\tau}\right)_\text{modulation} = \sqrt{\frac{D_{\text{eff},x}}{2c}} \times \frac{V_m}{D_x} \, .
\end{align}

Weak scattering can be approximated by adding a single very bright image to a screen. If the images move fast enough through the centre of the screen that the central image changes within an observation, this might be approximated by an image whose location is not fixed but determined by the shortest optical path through that screen, ignoring the dispersive phase structure. Hence, the stationary phase condition with respect to the screen that is considered weak needs to be applied on \cref{Eq:2scr_phitilde} with respect to the parallel coordinate $x_\shortparallel$ or $y_\shortparallel$. The coordinate of the new central image if the first screen is weak is thus given by
\begin{align}
    X_\shortparallel =& \frac{D_y D_{x,s} - D_x D_{y,s}\delta^2}{D_s D_{y}} P_\shortparallel + \frac{D_x}{D_y}\delta Y_\shortparallel \nonumber\\
    &+ \frac{D_x D_{y,s}}{D_s D_y}\gamma\delta P_\perp + \frac{D_x}{D_s}\gamma S_\perp
    \label{Eq:CPx}
\end{align}
and in the case of the second screen being weak, this image has to be added:
\begin{align}
    Y_\shortparallel =& \frac{D_y D_{x,s} - D_x D_{y,s}\delta^2}{D_s D_{x,s}} S_\shortparallel + \frac{D_{y,s}}{D_{x,s}}\delta X_\shortparallel \nonumber\\
    &- \frac{D_{y,s}}{D_s}\gamma P_\perp - \frac{D_x D_{y,s}}{D_s D_{x,s}}\gamma\delta S_\perp \, .
    \label{Eq:CPy}
\end{align}
Note that these images vary in position depending on the currently regarded image on the other screen. If both screens contain a central image with nonzero amplitude, the interaction of these images happens at locations that resemble the direct line of sight to the pulsar. The same result can be obtained by employing the intercept theorem or by solving \cref{Eq:CPx,Eq:CPy} simultaneously:
\begin{align}
    X_\shortparallel &= \frac{D_{x,s}}{D_s}P_\shortparallel + \frac{D_x}{D_s}\left( \delta S_\shortparallel + \gamma S_\perp \right) \, , \\
    Y_\shortparallel &= \frac{D_{y}}{D_s}S_\shortparallel + \frac{D_{y,s}}{D_s}\left( \delta P_\shortparallel - \gamma P_\perp \right) \, .
\end{align}




\bsp	
\label{lastpage}
\end{document}